%% file: AApaper.tex
\documentclass[traditabstract]{aa}

\usepackage{graphicx}
\usepackage{txfonts}

\usepackage{supertabular}
\usepackage{amsbsy}
\usepackage{natbib}
\bibpunct{(}{)}{;}{a}{}{,}

\usepackage{color}
\newcommand{\tmp}[1]{#1}
\newcommand{\tmpr}[1]{#1}
\newcommand{\refere}[1]{#1}
\newcommand{\conan}[1]{#1}
\newcommand{\vtk}[1]{}
\newcommand{\hide}[1]{}
\newcommand{\bdelta}{\mbox{\boldmath $\delta$}}
\newcommand{\bsigma}{\mbox{\boldmath $\sigma$}}
\newcommand{\bhdelta}{\mbox{\boldmath $\hat{\delta}$}}
\newcommand{\bhsigma}{\mbox{\boldmath $\hat{\sigma}$}}

\begin{document}

\title{Comparison of fringe-tracking algorithms for single-mode near-infrared long baseline interferometers}
\author{\'E. Choquet\inst{1,2,3} \thanks{\email{choquet@stsci.edu}} 
	\and J. Menu \inst{4,}\thanks{\tmp{PhD Fellow of the Research Foundation -- Flanders (FWO)}}
	\and G. Perrin\inst{1,2}
	\and F. Cassaing\inst{2,5}
	\and S. Lacour\inst{1,2}
	\and F. Eisenhauer\inst{6}
}
\institute{LESIA, Observatoire de Paris, CNRS, UPMC, Universit\'e Paris-Diderot, Paris Sciences et Lettres Research University, 5 place Jules Janssen, 92\,195 Meudon, France
	  \and Groupement d'Int\'er\^et Scientifique PHASE (Partenariat Haute r\'esolution Angulaire Sol Espace) between ONERA, Observatoire de Paris, CNRS and Universit\'e Paris Diderot
	 \and Space Telescope Science Institute, 3700 San Martin Drive, Baltimore MD 21218, USA
	  \and Instituut voor Sterrenkunde, KU Leuven, Celestijnenlaan 200D, 3001 Leuven, Belgium 
	  \and \tmp{Onera -- The French Aerospace Lab} , BP 72, 92\,322 Ch\^atillon, France
	  \and Max Planck Institute for extraterrestrial Physics, PO Box 1312, Giessenbachstr., 85\,741 Garching, Germany
}
\date{Received dd Month year/
      Accepted dd Month year}

\abstract{To enable optical long baseline interferometry toward faint objects, long integrations are necessary despite atmospheric turbulence. Fringe trackers are needed to stabilize the fringes and thus increase the fringe visibility and phase signal-to-noise ratio (S/N), with efficient controllers robust to instrumental vibrations and to subsequent path fluctuations and flux drop-outs.

We report on simulations, analysis, and comparison of the performances of a classical integrator controller and of a Kalman controller, both optimized to track fringes \refere{under realistic observing conditions for different source magnitudes, disturbance conditions, and sampling frequencies. The key parameters of our simulations (instrument photometric performance, detection noise, turbulence, and vibrations statistics) are based on typical observing conditions at the Very Large Telescope observatory and on the design of the GRAVITY instrument, a 4-telescope single-mode long-baseline interferometer in the near-infrared, next in line to be installed at VLT Interferometer.}

We find that both controller performances follow a two-regime law with the star magnitude, a constant disturbance limited regime, and a diverging detector- and photon-noise-limited regime. Moreover, we find that the Kalman controller is optimal in the high and medium S/N regime owing to its \refere{predictive commands} based on an accurate disturbance model. In the low S/N regime, the model is not accurate enough to be more robust than an integrator controller. Identifying the disturbances from high S/N measurements improves the Kalman performances in the case of strong optical path difference disturbances.}

\keywords{Instrumentation: high angular resolution -- 
	  Atmospheric effects -- 
	  Methods: numerical -- 
	  Techniques: interferometric}

\maketitle
\section{Introduction\label{Intro}}

Atmospheric turbulence is a major limiting factor for the sensitivity of ground-based optical long-baseline interferometers. Because of the short coherence time of atmospheric turbulence -- typically $\tau_0=20$~ms in the near infrared -- basic observations are indeed limited to short exposure times and, as a consequence, to bright targets. \refere{With long exposures}, the random fringe motion induced by the turbulent atmosphere would 
 completely blur fringe contrast and \refere{would provide poor phase and visibility measurements.}

To overcome this limitation, a servo-system called fringe tracking has been developed by \citet{Shao1977} and demonstrated three years later \citep{Shao1980}. By stabilizing the interferometric fringes to a fraction of the wavelength, such a system indeed enables exposure times of several minutes without the visibility loss inherent to the fringe motion, and therefore gives access to new sensitivity limits. Fringe trackers have since proven themselves to be essential, not only to observe faint targets \refere{\citep{Muller2010,Colavita2013}}, but also to perform astrometry with a sub milliarcsecond (mas) accuracy \refere{\citep{Lane2004}}.
That explains why the main long baseline interferometers are currently being provided either with instruments including their own fringe trackers: PRIMA \citep{delplancke2006,Sahlmann2009}, GRAVITY \citep{eisenhauer2011}, or with shared
fringe trackers: fringe tracker of the Keck Interferometer \citep{Colavita2010}, CHAMP for the CHARA array \citep{Berger2008}, FINITO for the Very Large Telescope Interferometer (VLTI) \citep{LeBouquin2008}.

\refere{Another key parameter to reach high sensitivity limits with interferometer is to combine telescopes with large apertures, to obtain a large collecting surface. However, most of the 10~m-class telescopes suffer from strong vibrations, due to their lightened structure more subject to have vibrating frequencies excited either by the wind, telescopes movement while tracking a star, or by the instruments fixed at their different focuses which are usually equipped with highly vibrating systems (electronics, coolers\ldots). Stabilizing fringes in such conditions is particularly challenging because vibrations lead to important path length fluctuations which add up to the atmospheric turbulence, and also generate tip-tilt variations of the beams, resulting in fluctuations of the combined flux. A classical controller is typically unable to correct for system vibrations because of its continuous transfer function with limited bandwidth. At best, a classical controller can dampen the lowest-frequency components which are well-within its bandwidth, but it can not completely filter out disturbances at a specific frequency.}

\refere{To overcome this problem, several approaches can be combined. First of all, the sources of the vibration excitations can sometimes be identified and switched off. An efficient example of such efforts have been demonstrated at VLTI these past years, bringing the optical path length variations on the 8.2~m unit telescopes (UTs) from typical values of 280 nm rms to more than 1~$\mu$m rms in 2008 down to 145 to 380~nm rms in 2012, by damping pumps of some instruments, changing the close cyclo-coolers of some others, and modifying the mechanical design of some mounts \citep{Poupar2010,Haguenauer2010,Haguenauer2012}. Secondly, vibrations excited by the first mirrors of the system can be measured with accelerometers and independently compensated with optical delay lines \citep{Colavita2013,Haguenauer2012}. However, these solutions are not enough to completely suppress vibrations down the entire optical stream, and active compensation is necessary at the beam combination level. If a handful of vibrations are properly identified and characterized by a phase sensor, individual narrow-band suppression control blocks can be used to filter them out. Such a solution was implemented at Keck interferometer with higher-harmonic controllers (HHC) \citep{Colavita2010}, and at VLTI with an adaptive version, the vibration-tracking algorithm (VTK) \citep{Dilieto2008}.}

An alternative approach is to compensate all vibrations \refere{in the optical path \emph{and} the atmospheric} turbulence at the same time. This can be \refere{achieved with} a Kalman controller, which is a predictive algorithm \refere{whose commands are} based on a model of the \refere{identified disturbance components}. The formalism was first developed by \citet{Kalman1960}, then has been transposed to adaptive optics (AO) systems by \citet{LeRoux2004} and \citet{Petit2004}. \conan{First on-sky demonstration of full wavefront control with a Kalman filter has been achieved in 2012 and 2013 on CANARY MOAO pathfinder \citep{Sivo2013}, and Kalman-based tip-tilt correction has been implemented in two extrem AO systems: the Gemini Planet Imager (GPI) at Gemini south \citep{Hartung2013} and the Spectro-Polarimetric High-contrast Exoplanet Research (SPHERE) instrument at VLT \citep{Petit2012}. First light of these two instruments is planned in 2014}. Based on these developments in wavefront control, Kalman filtering was adapted to 2-way fringe tracking by \citet{Lozi2011} and to 4-way fringe tracking by \citet{Menu2012}. 

\refere{In this paper, we aim} at comparing the performances of a Kalman controller and a \refere{classical} controller, \refere{both optimized to track fringes in different disturbance conditions with multiple baselines. We performed numerical simulations to analyze their robustness against several levels of instrumental vibrations and to coherent flux variations.} 
\refere{To do so, we based our simulations on the design of the GRAVITY instrument, which is the next instrument in line to be installed at VLTI, and provides thus the perfect framework to compare control algorithms for fringe tracking. }

\refere{Currently,} the VLTI is the most powerful observatory to match high sensitivity and high angular resolution. \refere{On the one hand,} it is indeed the only long-baseline interferometer allowing the combination of four 8.2~m UTs with thus the greatest available collecting surface \refere{to date}, \refere{on 47~m to 130~m baselines,} and \refere{on the other hand, it has the capacity to combine four relocatable 1.8~m auxiliary telescopes (ATs) on} baselines with lengths ranging from 8~m to 202~m \refere{(only 11~m to 140~m baselines are offered at present)}, with thus a \refere{potential} resolution of 2~mas at 2~$\mu$m. \refere{However, each UT is equipped with several instruments fixed at its different Cassegrain focuses which excite vibration frequencies differing from one UT to another. Despite strong effort to minimize them, these vibrations still limit operations at VLTI \citep{Merand2012}.} 

\refere{GRAVITY (General Relativity Analysis via VLT InTerferometrY) is a second-generation instrument for the VLTI that} will combine up to four telescopes in the K band, and is currently in phase D with first on-sky tests in 2015 \citep{eisenhauer2011}. \refere{Its goal is to provide dual-field precision astrometry of order of 10~$\mu$as and phase-referenced imaging with 4~mas resolution, with as primary science case the study} of the close environment of Sgr A*, the supermassive black hole in the center of the Galaxy. \refere{With this objective}, it will lead to unprecedented imaging of targets of magnitudes up to $K=16$ with 100~s exposure time \citep{gillessen2010}, thanks to fringe tracking on an unresolved reference star of magnitude up to 10, which is the current best sensitivity limit in this band \citep[Keck Interferometer,][]{Ragland2010}. \refere{To reach this goal, the fringe tracker will have to stabilize the optical path differences (OPDs) down to 350~nm rms on six interferometric baselines with the UTs under median atmospheric conditions, using a controller robust to longitudinal vibrations and coherent flux variations.}

\refere{We present in this paper realistic numerical simulations of fringe tracking under different disturbance conditions based on the design of GRAVITY, and compare the performances of a Kalman controller and of an integrator controller in this framework.} In Sect.~\ref{simus}, we describe the simulation of the \refere{disturbances}, and in Sect.~\ref{simus_detec} the parameters used \refere{to simulate the combining and detection optical systems}. In Sect.~\ref{algo} we present the algorithms used to estimate the OPD on each baseline, \refere{and we detail both controllers}. In Sect. \ref{results}, \refere{we describe the results of the simulations and compare the} performances of the integrator and Kalman controllers, \refere{for different observing conditions and reference star magnitudes}. These results are discussed in Sect.~\ref{discussion}, \refere{and comments are pointed out for the specific case of GRAVITY. Finally, the conclusions of this paper are summarized in Sect.~\ref{conc}.}

\section{\tmp{Simulation of the disturbances} \label{simus}}

We analyze the closed-loop behavior of the fringe tracker with an iterative time-domain simulator\footnote{\refere{Developed in the Interactive Data Language (IDL) programming language.}} reproducing the entire acquisition and control process of the fringe tracker. Discrete sequences of disturbance 
are \refere{generated as input of} the simulator. At each iteration, the image on the fringe tracker detector is computed from the OPD residuals, and is then processed by the algorithm of the fringe tracker to deduce new piston commands.

\refere{The parameters that we used to simulate the disturbances and the beam combination are based on the median atmospheric conditions at the VLT observatory at Cerro Paranal, Chile, and on the design of the fringe tracker of the GRAVITY instrument.}
\refere{These parameters} are listed in Table~\ref{fixed parameters}. 
\refere{Moreover,} we simulated different observing conditions to evaluate the robustness of the controllers. The parameters and their range of variations are listed in Table~\ref{var param}. \refere{Table~\ref{spec} indicates the specifications of the fringe tracker of GRAVITY.}

\refere{We describe in} this section the \refere{models used to} simulate the disturbances, a.k.a. the atmospheric piston, longitudinal vibrations, and flux variations.

\begin{table}
 \caption{\refere{Fix parameters used in the simulations}.} 
 \label{fixed parameters} 
 \centering 
 \begin{tabular}{c c} 
  \hline\hline 
  Item & Value\\
  \hline
  Average wavelength ($\mu$m)& 2.2\\
  Spectral bandwidth ($\mu$m) & 0.5\\
  Number of spectral channels & 5\\
  Number of telescopes & 4 \\
  Number of baselines & 6 \\
  Telescope diameter (m)& 8.2 \\
  Average baseline length (m)& 80\\
  Total transmission\tablefootmark{1} (\%)& \refere{1.0} \\
  Detector RON ($\mathrm{e}^{-}$)& 4\\
  APD\tablefootmark{2} noise excess factor & 1.5\\
  Seeing (\arcsec)& 0.8 \\
  Atmospheric coherence time (ms)& 3.4 \\
  Wind speed (m/s)& 12\\
  Atmospheric turbulence outer scale (m)& 100 \\
  Number of frames & 30\,000\\
  \hline
 \end{tabular} \begin{flushleft}\tablefoottext{1}{Transmission from primary to detector, with fiber coupling efficiency excluded.}
\tablefoottext{2}{Avalanche Photo-Diode.}      
 \end{flushleft}
\end{table}

\begin{table}
 \caption{\refere{Varying parameters and their range of variation.}}
 \label{var param} 
 \centering 
 \begin{tabular}{c c} 
  \hline\hline 
  Item & Values\\
  \hline
  \tmp{K-band} star magnitude & 6 -- 12\\
  Tilt per axis (mas rms)& 15 \& 20\\
  Vibration levels & \refere{Null; Low; High}\tablefootmark{1}\\
  Atmospheric OPD ($\mu$m rms) & 10 \& 15\\
  Loop frequency (Hz) & 100 -- 1000\\
  Kalman model (frames)& 2\,000 -- 5\,000\\
  \hline
 \end{tabular}
 \begin{flushleft} \tablefoottext{1}{See Table~\ref{vibs_rms} for the \refere{related} RMS values \refere{per telescope}.}      
 \end{flushleft}

\end{table}

\begin{table}
 \caption{\refere{Specifications of the fringe tracker of GRAVITY.}}
 \label{spec} 
 \centering 
 \begin{tabular}{c c} 
  \hline\hline 
  Item & Values\\
\hline
 \multicolumn{2}{c}{\emph{Conditions}} \\
\hline
Reference magnitude&10\\
Seeing (\arcsec)&0.8\\
Vibration level (nm rms)&150\\
Residual tip-tilt (mas rms)&15\\
  \hline
 \multicolumn{2}{c}{\emph{Goal}} \\
\hline
OPD residuals (nm rms) & $\le 350$\\
  \hline
 \end{tabular}
\end{table}

\subsection{Modeling of piston disturbances}
\refere{As first input of our simulator, we generate a sequence of piston fluctuations over $N$ frames} for each aperture. The sequence of piston disturbances, hereafter $\{\vec{P}_n\}_{n\in\llbracket1,N\rrbracket}$ \refere{where $\vec{P}_n$ is a four-element vector in our GRAVITY-like case}, is computed from the sum of the two main causes of disturbance: the atmospheric piston and the longitudinal vibrations.

\subsubsection{Atmospheric piston \label{Atm_dist}}

To simulate the atmospheric turbulence on each aperture, we used a Von K\'arm\'an model \citep{Reinhardt1972}, with a refractive-index spatial power spectrum $\Phi_\mathrm{N}$ proportional to
\begin{equation}
 \Phi_\mathrm{N}(\kappa) \propto \left( \kappa^2+1/L_0^2 \right)^{-11/6},
\end{equation}
with $\kappa$ the spatial frequency. Unlike a basic Kolmogorov model which diverges at zero frequency, this model includes a finite atmospheric outer scale $L_0$, which leads to the saturation of the spectrum at low frequencies \refere{and is thus more realistic}. 

The asymptotic OPD temporal power spectrum density (PSD) $S_\mathrm{atm}$ for a Von K\'arm\'an model is described by \citet{Buscher1995}. \refere{For given wind speed $V$ and baseline length $B$,} it follows an $f^{0}$ law for low temporal frequencies $f$, an $f^{-2/3}$ power law between the two cut-off frequencies $f_1 \sim 0.2V/B$ and $f_2 \sim V/L_0$, and an $f^{-8/3}$ law for higher frequencies:
\begin{equation}
 S_\mathrm{atm}(f) \propto \left \{ \begin{array}{ll}
					    f^{0} & \mathrm{if}\quad f<f_1 \\
					    f^{-2/3} & \mathrm{if}\quad f_1<f<f_2 \\
					    f^{-8/3} & \mathrm{if}\quad f>f_2 \\
					    \end{array}
				  \right. .
\end{equation}
Atmospheric piston sequences are simulated \refere{for each aperture} by multiplying in Fourier space the spectrum of a white Gaussian noise sequence with this asymptotic power spectrum. The time-sequences are then scaled to match \refere{a given atmospheric OPD standard deviation} $\sigma_\mathrm{atm}$ \refere{when combining two apertures}.

\begin{figure}
 \resizebox{\hsize}{!}{\includegraphics{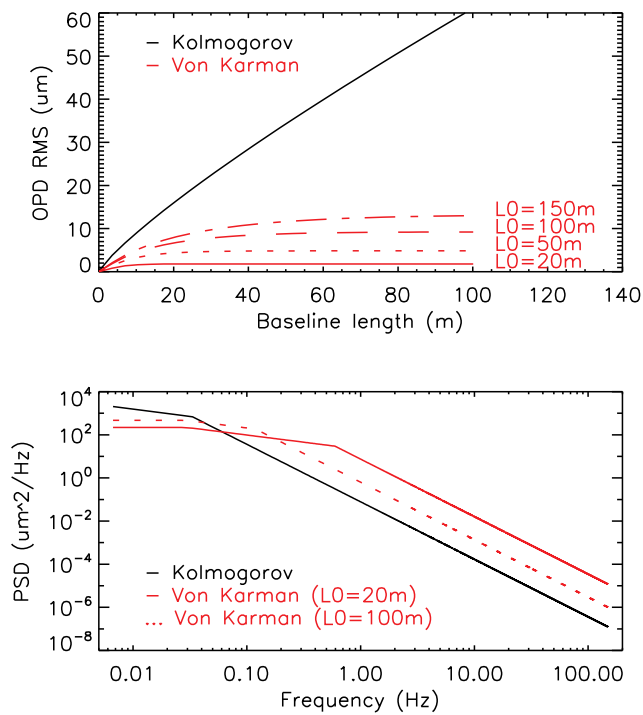}}
 \caption{\refere{Comparison of Kolmogorov and Von K\'arm\'an models with different atmospheric outer scale values. \textbf{Top}: standard deviation of OPD fluctuations as a function of the baseline length for median atmospheric conditions in the K band \citep{Conan2000,Maire2006}. \textbf{Bottom}: OPD temporal PSDs, scaled to have identical energy level (standard deviation of 10~$\mu$m).}}
 \label{outerscale}
\end{figure}

\begin{figure}
 \resizebox{\hsize}{!}{\includegraphics{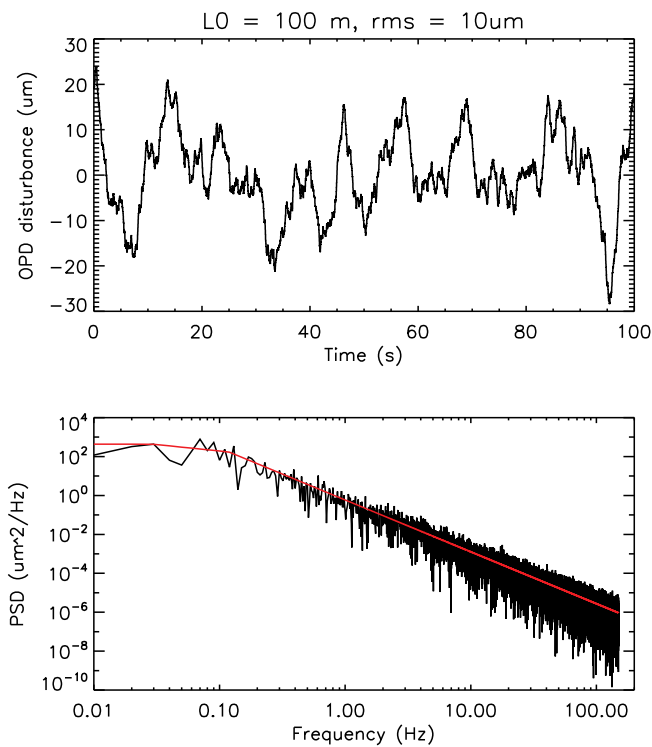}}
 \caption{\refere{Typical simulated atmospheric OPD disturbance, with a Von K\'arm\'an model, an atmospheric outer $L_0$ of 100m, and a standard deviation of 10~$\mu$m. Top: temporal sequence. Bottom: simulated PSD (\textit{black}) and PSD model used for the simulation (\textit{red}).}}
 \label{opd_atm}
\end{figure}

\refere{A typical value of $L_0=22$~m was measured at Paranal with the Generalized Seeing Monitor (GSM) instrument in 1998 and 2007 \citep{Martin2000,DaliAli2010}. } 
\refere{However, we do not use this value in our simulations for two main reasons. First of all, the standard deviation predicted by \citet{Conan2000} and \citet{Maire2006} for the piston fluctuations with an atmospheric outer scale of 22~m is much lower than the typical value of 10~$\mu$m measured under median seeing  conditions in the K band with PRIMA \citep{Sahlmann2009} or in the H band with FINITO \citep{LeBouquin2008}. The standard deviation of OPD fluctuations for different values of $L_0$ in the K band are illustrated in Fig.~\ref{outerscale} (top). Furthermore, at identical energy level, the atmospheric piston spectrum with a Von K\'arm\'an model and a low outer scale value have stronger high-frequency components, which are not representative from OPD spectrum measured at VLTI with PRIMA and FINITO (cf. Fig.~\ref{outerscale} bottom).} 
\refere{This important difference with the GSM measurements is attributed to an instrumental contribution from the VLTI in the OPD fluctuations (e.g. internal turbulence in the delay lines).}
\refere{In order to generate realistic disturbance sequences similar to OPD measurements at VLTI, we thus adopted a value of 100~m for the atmospheric outer scale, and scaled the atmospheric disturbance to a typical standard deviation of $\sigma_\mathrm{atm}=10~\mu$m. In Fig.~\ref{opd_atm}, we present a typical atmospheric OPD temporal sequence and PSD used for these simulations.}

\subsubsection{Piston vibrations}
\refere{To this atmospheric turbulence sequence, we add a discrete number of vibrations for each aperture} to simulate the total piston disturbance sequence $\{\vec{P}_n\}_{n\in\llbracket1,N\rrbracket}$. 
Strong vibrations are indeed excited by the instruments installed at the focuses of the \refere{telescopes} and propagate along their \refere{mechanical} structure. We modeled each \refere{narrow-band} vibration peak by a damped harmonic oscillator excited by a Gaussian noise $v$, characterized by three parameters: the natural frequency $f_\mathrm{0}$, the damping coefficient $k$, and the variance of the excitation $\sigma_\mathrm{v}^2$. The temporal evolution $p_\mathrm{v}(t)$ of a piston vibration is defined by the following differential equation \citep{Petit2006}:
\begin{equation}
 \frac{\mathrm{d}^2p_\mathrm{v}}{\mathrm{d}t^2}+4\pi{}kf_0\frac{\mathrm{d}p_\mathrm{v}}{\mathrm{d}t}+4\pi{}^2f_0^2p_\mathrm{v}(t)=\frac{v(t)}{T^2},
\end{equation}
where $T$ is the sampling pitch. The related temporal PSD $S_\mathrm{vib}$ is thus expressed by
\begin{equation}
 S_\mathrm{vib}(f)=\frac{\sigma_\mathrm{v}^2T/(16\pi^4T^4)}{f^4+2f_\mathrm{0}^2f^2(2k^2-1)+f_\mathrm{0}^4}.  \label{eq_vibr}
\end{equation}
\refere{Each vibration is simulated by} multiplying in the Fourier space the model PSD with the spectrum of a white Gaussian noise sequence. \refere{We assume in our simulations that the vibrations have invariant characteristics. This point is further discussed in Sect.~\ref{discussion}.}

\begin{figure}
 \resizebox{\hsize}{!}{\includegraphics{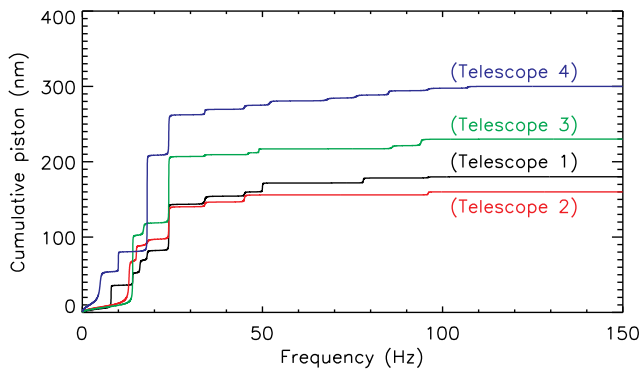}}
 \caption{Cumulative piston modeled for each telescope \refere{for the high vibration level}.}
 \label{Cumu}
\end{figure}

\refere{Three} different levels of vibrations are tested in our simulations: \refere{no vibration at all, a low vibration level as expected in late 2014 at VLTI, and a high vibration level similar to the current level on the UTs}. For the current conditions \refere{at VLTI}, we used the vibration parameters listed in Table~\ref{vibs} for the four telescopes. \refere{These values have been chosen to reproduce similar cumulative piston vibrations and vibration spectra measured on the UTs by \citet{Dilieto2008} and \citet{Poupar2010}. The adopted damping coefficients correspond to resonance of 54~dB for the narrowest vibrations ($k=0.001$), and 20~dB for the more damped ones ($k=0.05$). We used the same vibration parameters for the low vibration level}, and scaled the total OPD standard deviations to $\sigma_\mathrm{vib}=150~$nm \refere{on each baseline}. \refere{We provide a summary of} the total standard deviations of the vibrations in Table~\ref{vibs_rms} for the three simulated levels. \refere{In Fig.~\ref{Cumu}, we present} the cumulative vibration spectra \refere{per telescope simulated for the high vibration level.}

\begin{figure}
 \resizebox{\hsize}{!}{\includegraphics{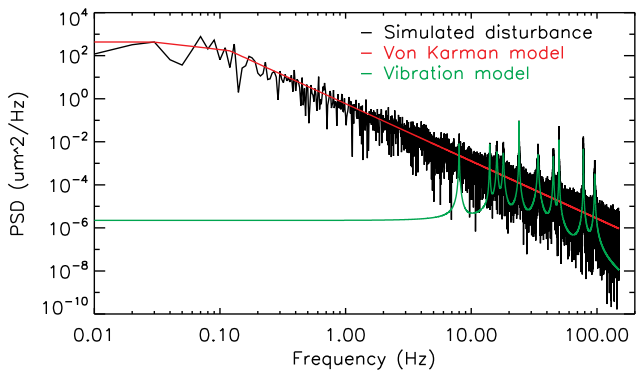}}
 \caption{PSD of a disturbance sequence on telescope 1 (\textit{black}) and models used \refere{for the simulation}: Von K\'arm\'an model for the atmospheric piston (\textit{red}), and vibration model \refere{for the high vibration level} (\textit{green}).}
 \label{PSD_vib}
\end{figure}

The PSD of a disturbance sequence simulated for telescope~1 is shown in Fig.~\ref{PSD_vib}, with the models for both the atmospheric turbulence and the longitudinal vibrations, \refere{when the telescope is subject to the high level of vibration}.

\begin{table}
 \caption{Parameters of the vibration peaks simulated on each telescope. The \refere{parameter $\sigma_v$  is given for the high level of vibrations.}}
 \label{vibs} 
 \centering 
 \begin{tabular}{c c c c c c c} 
  \hline\hline 
  \multicolumn{3}{c}{Telescope 1} && \multicolumn{3}{c}{Telescope 2}\\
  $f_\mathrm{0}$ (Hz) & $k$ & $\sigma_v$ (nm) && $f_\mathrm{0}$ (Hz) & $k$ & $\sigma_v$ (nm)\\ 
  \hline 
  8  & 0.003 & 0.25	&&	13  & 0.01  & 1.8 \\
  14 & 0.002 & 0.5 	&&	15  & 0.003 & 1.0  \\
  16  & 0.006 & 1.3 	&&	18  & 0.02  & 2.5   \\
  18  & 0.006 & 1.5	&&	24  & 0.002 & 3.0  \\
  24  & 0.001 & 2.5 	&&	34  & 0.004 & 3.0  \\
  34  & 0.006 & 5.0 	&&	45  & 0.003 & 5.0  \\
  45 & 0.003 & 4.0   	&&	96  & 0.001 & 6.0\\
  50 & 0.001 & 4.0	&&	&&\\
  78 & 0.001 & 6.0	&&	&&\\
  96 & 0.003 & 7.0	&&	&&\\

  \hline \hline 
  \multicolumn{3}{c}{Telescope 3} & \multicolumn{3}{c}{Telescope 4}\\
  $f_\mathrm{0}$ (Hz) & $k$ & $\sigma_v$ (nm) && $f_\mathrm{0}$ (Hz) & $k$ & $\sigma_v$ (nm)\\ 
  \hline
  14  & 0.002 &1.4 	&&	5  & 0.05& 0.8  \\
  17  & 0.01  &2.5 	&&	10 & 0.002 & 0.5    \\
  24  & 0.001 &3.7  	&&	18 & 0.001 & 2.8   \\
  34  & 0.003 &2.0  	&&	24 & 0.002 & 5.0   \\
  46  & 0.002 &2.7  	&&	34 & 0.003 & 4.0    \\
  49  & 0.001 &3.0  	&&	45 & 0.004 & 6.2  \\
  86  &	0.003 &11.0 	&&	52 & 0.005 & 9.0    \\
  94  &	0.002 &15.0	&&	68 & 0.007 & 13.0  \\ 
  &&			&&	76 & 0.006 & 15.0\\
  &&			&&	85 & 0.002 & 12.0\\
  &&			&&	96 & 0.005 & 18.0\\
  &&			&&	107& 0.002 & 11.0\\
 \end{tabular}
\end{table}

\begin{table}
 \caption{\refere{Total} standard deviation of \refere{vibrations per telescope, for each simulated level}.} 
 \label{vibs_rms} 
 \centering 
 \begin{tabular}{c c c c c} 
  \hline\hline 
  Vibration & \multicolumn{4}{c}{Total vibration RMS (nm)}\\
  level & Tel. 1 & Tel. 2 & Tel. 3 & Tel. 4\\
  \hline
  \refere{Null} & 0 & 0 & 0  & 0  \\
  \refere{Low} & 106 & 106 & 106 & 106\\
  \refere{High} & 180 & 160 & 230 & 300\\
  \hline
 \end{tabular}
\end{table}

\subsection{Flux variations}
\refere{When working close to the sensitivity limit of the fringe tracker, flux variations in the beams  induce serious performance losses: inaccurate commands might be send to the actuators, or commands might be frozen as long as the precision on the estimated OPDs is too low.}

\refere{The second input of our simulator is a temporal sequence of flux variations in the beam combiner, hereafter $\{\vec{F}_n\}_{n\in\llbracket1,N\rrbracket}$ where $\vec{F}_n$ is the four-element vector of the number of photon in each combined beam.} For a single-mode fiber interferometer like GRAVITY, the flux variations come from residual wavefront errors and tip-tilt fluctuations of the beams at the input of the fibers.

\subsubsection{Total coherent flux}
The \tmp{brightness} $E$ of a star of magnitude $K$ is defined by
\begin{equation}
 E=E_0 10^{-K/2.5},
\end{equation}
with $E_0=670$~Jy in the K band \citep{campins1985}.

The maximum coherent flux $F_\mathrm{max}$ coming to the input of each fiber during one frame is computed for the whole K band by
\begin{equation}
 F_\mathrm{max}=T\frac{\pi D^2}{4R_{bb}f_\mathrm{FT}}E,
\end{equation}
with $D$ the diameter of the telescope, $R_{bb}$ the broad-band spectral resolution in the K band, $f_\mathrm{FT}$ the frame rate, and $T$ the overall transmission of the system, from the primary mirror to the detector \refere{(i.e. beam combiner transmission and detector quantum efficiency included)}.

We simulated an overall transmission of 1~\% as \refere{expected with the GRAVITY instrument and the VLTI facility}. This leads to a typical flux of 400~photons per aperture and per frame at the input of the beam combiner for a magnitude K=10 star, a telescope diameter of 8.2~m, and a loop frequency of 300~Hz.

\subsubsection{Fiber injection efficiency}

 The tip-tilt fluctuations leading to flux variations come from three different contributions: \refere{telescope} vibrations, residual wavefront aberrations from the AO, and tilt errors from the beam relay and guiding system.

We modeled the tilt vibrations on the \refere{telescopes} by a pure sinusoidal vibration at frequency 18.1~Hz with random phases for each telescope, and a standard deviation of 5~mas. \refere{This vibration is indeed the dominant one observed in tip-tilt on VLT/NACO}.

We modeled the residual tilt from the AO system as well as the tilt from guiding errors with the temporal PSD measured with the IRIS tilt sensor \citep{Gitton2004, Gitton2008}. We used the following empirical model for the spectrum $S_\mathrm{IRIS}$, deduced from past ESO measurements\footnote{Philippe Gitton private communication}:
\tmp{\begin{equation}
 S_\mathrm{IRIS}(f) \propto \left \{ \begin{array}{ll}
				  \log(f/f_1)/\log(f_2/f_1)  & \mathrm{if}\quad f_1<f<f_2 \\
				  \log(f/f_3)/\log(f_2/f_3) & \mathrm{if}\quad f_2<f<f_3 \\
				  0 & \mathrm{otherwise} \\
				\end{array}
			  \right. ,
\end{equation}}
with the cut-off frequencies $f_1=2$~Hz, $f_2=8$~Hz and $f_3=50$~Hz. We multiplied this model with the spectra of white Gaussian noise sequences to get random tilt sequences \refere{which are then scaled to standard deviations of 8.8~mas and 10.5~mas for the AO residuals and the guiding errors, respectively. Such values can be achieved at VLTI with a dedicated guiding camera which would stabilize the beams in the VLTI lab, as is planned for GRAVITY.}

\begin{figure}
 \resizebox{\hsize}{!}{\includegraphics{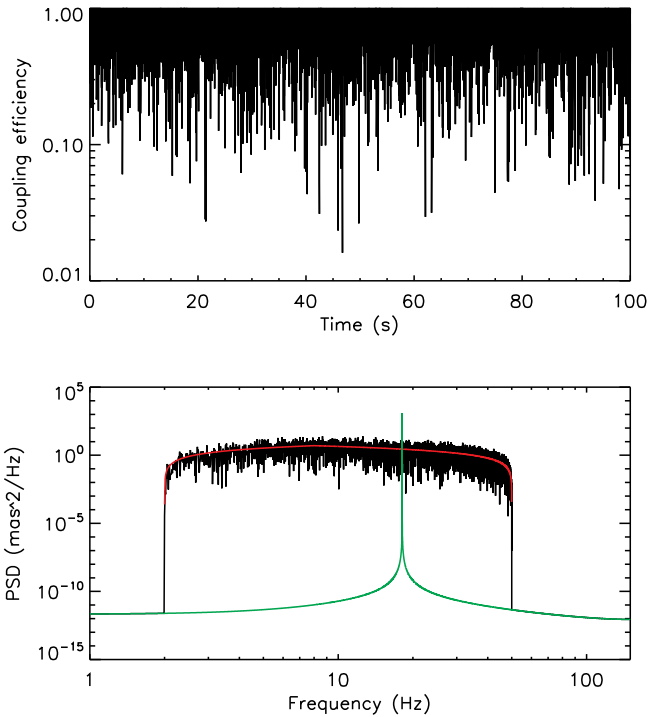}}
 \caption{\refere{Top:} simulated coupling efficiency \refere{temporal} sequence. \refere{Bottom: PSD of a tilt sequence (\textit{black}), vibration added (\textit{green}), and model used for AO and Guiding residuals (\textit{red}).}}
 \label{PSD_IRIS}
\end{figure}

The total tip-tilt sequences $\{\vec{\theta}_n\}_{n\in\llbracket1,N\rrbracket}$ are computed from the sum of these three simulated tilt sequences.

To compute the coupling ratio into the single-mode fibers, we assumed that the fiber coupler numerical aperture is optimized for the mode of the fibers. Following \citet{Wallner2002}, the fiber mode field radius $\omega_0$ is then
\begin{equation}
 \omega_0\simeq0.714\frac{\lambda_0f_\mathrm{c}}{D},
\end{equation}
with $f_\mathrm{c}$ the fiber coupler focal length. The coupling efficiency $\eta$ for a beam tilted by an angle $\theta$ with respect to the optical axis is:
\begin{equation}
 \eta(\theta)= \eta_0 \exp\left( - 2\left(\frac{f_c\theta}{\omega_0}\right)^2\right),
\end{equation}
with an optimum coupling efficiency $\eta_0=81\%$ (without tilts and misalignments).

We thus simulated the sequence $\{\vec{F}_n\}_{n\in\llbracket1,N\rrbracket}$ of coherent flux fluctuations such that:
\begin{equation}
 \vec{F}_n=F_\mathrm{max}\tmp{\eta_0} \exp\left( - 2\left(\frac{f_c\vec{\theta}_n}{\omega_0}\right)^2\right).
\end{equation}

For a telescope of diameter $D=8.2$~m, the average loss from optimal coupling is 80~\% with a standard deviation of 20~\%. A typical normalized flux sequence \tmpr{is} presented in Fig.~\ref{PSD_IRIS}.

\section{\tmp{Simulation of the detection system}\label{simus_detec}}

\subsection{Acquisition sequence}
\begin{figure}
 \resizebox{\hsize}{!}{\includegraphics{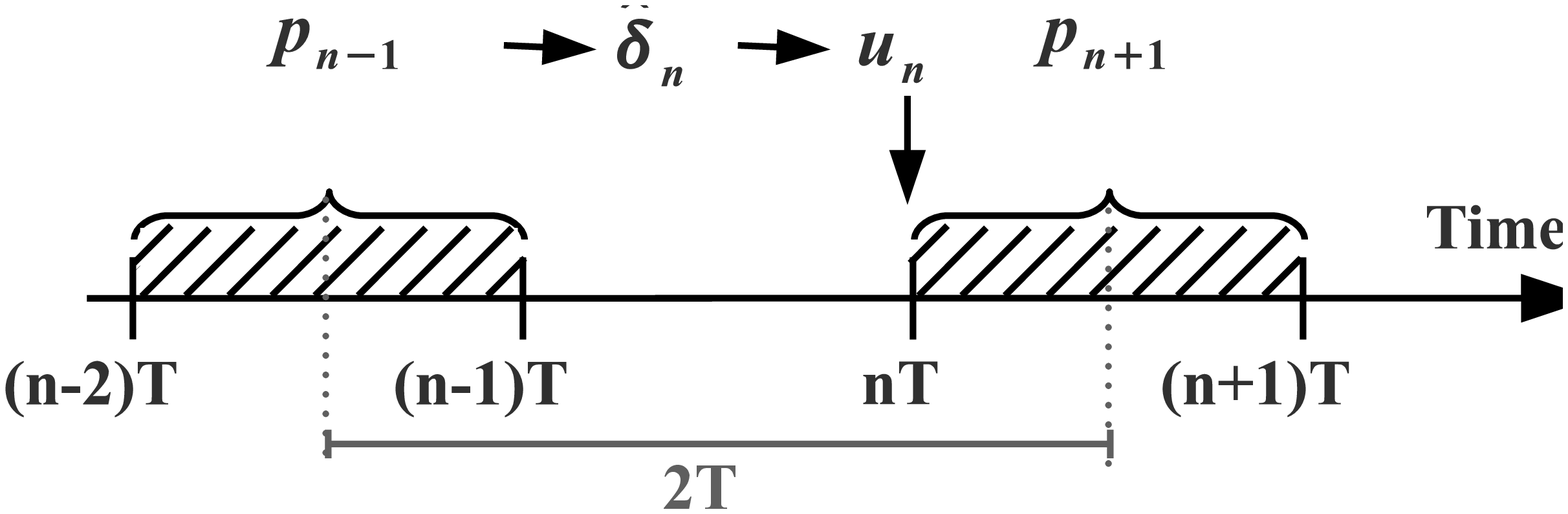}}
 \caption{Representation of the simulated discretized time-scheme. At iteration $n$, the OPDs $\vec{\hat{\delta}}_n$ are estimated from the image $\vec{I}_{n-1}$ delivered at the end of the previous iteration, which is representative of piston disturbances $\vec{P}_{n-1}$ integrated between time steps $n-2$ and $n-1$. The corresponding correction on the delay lines $\vec{U}_n$ is applied during the following iteration.}
 \label{Time_diag}
\end{figure}

To \refere{create} a realistic acquisition temporal process \refere{in our simulator}, we introduced a two-frame delay between the correction applied by the piston actuators and the corresponding effective disturbances. This takes into account one frame for the fringes to be acquired, and one frame for the fringe tracker to compute the corresponding corrections, which are applied at the beginning of the next frame. This time scheme is illustrated in Fig.~\ref{Time_diag}. The response of the actuators is assumed to be instantaneous at this point.

\subsection{Beam combiner and detector \label{BC}}

So as to properly estimate the residual OPDs (see Sect.~\ref{opd_est}), fringes \refere{are} spectrally dispersed \refere{to efficiently compute large OPD residuals. In our simulations,} the full K band is divided into $N_\lambda=5$ spectral channels with effective wavelengths $\left \{ \lambda_l\right \}_{\llbracket1,N_\lambda\rrbracket}=\left \{ 1.95~\mu\mathrm{m}, 2.075~\mu\mathrm{m}, 2.2~\mu\mathrm{m}, 2.325~\mu\mathrm{m}, 2.45~\mu\mathrm{m} \right \}$.

For each spectral channel $l$, the complex \refere{four-element} amplitude vector $\vec{A}_n^l$ is computed \refere{at iteration $n$} \refere{with} phases \refere{resulting} from the difference between the piston disturbances $\vec{P}_n$ and the piston actuator positions, i.e. the commands $\vec{U}_{n-1}$ computed at the previous iteration:
\begin{equation}
 \vec{A}_n^l=\sqrt{\frac{\vec{F}_n}{N_\lambda}}\exp\left(2\mathrm{i}\pi\frac{\vec{P}_n-\vec{U}_{n-1}}{\lambda_l}\right).
\end{equation}

To simulate the interferences, we used a pairwise ABCD beam combination with spatial coding \citep{Benisty2009}, where each couple of beams produces four intensities in phase quadrature (ABCD-like combination, \citet{Shao1977}). This configuration is indeed optimal to sample fringes from four telescopes as explained in \citet{Blind2011}, and corresponds to the fringe sampling chosen for GRAVITY. \refere{In actual} ABCD \refere{components}, the \refere{measured} modulations in phase opposition (A-C and B-D) are exactly $180^\circ$ due to energy conservation, whereas the quadrature phase shifts are non-perfect and weakly wavelength-dependent. 

\refere{We} simulated B-A phase shifts {representative} from measurements of the GRAVITY integrated optics beam combiner\footnote{Laurent Jocou private communication}, \refere{which} are detailed in Table~\ref{ABCD}. They present departures from the nominal $90^\circ$ phase-shift from $2^\circ$ to $17^\circ$ and an average chromatic spread of $9.8^\circ$. \refere{In addition, we simulated a} reduced contrast of 75~\% \refere{for each interference pattern,} to account for overall contrast loss of the instrument, partial polarization effects, and OPD turbulence residuals. 

\refere{The image generated by our simulator is a} $N_\lambda \times 24$ pixel dispersed image $\tens{I}_n$ where each row corresponds to one spectral channel. The \refere{24-element intensity vector} $\vec{I}_n^l$ at spectral channel $l$ is computed thanks to the corresponding visibility-to-pixel matrix (V2PM), following the formalism developed by \citet{Lacour2008}:
\begin{equation}
 \vec{I}_n^l=\tens{V2PM}^l \, \vec{C}_n^l,
\end{equation}
where $\tens{V2PM}^l$ is the $24\times 16$ V2PM matrix at spectral channel $l$, and $\vec{C}_n^l$ is the 16-element vector of the coherences, defined in three parts such that: 
\begin{equation}
 \vec{C}_n^l= \left[{\vec{C}_\mathrm{F}}_n^l \quad \Re \left({\vec{C}_\mathrm{C}}_n^l\right) \quad \Im \left({\vec{C}_\mathrm{C}}_n^l\right)\right]^{\rm T}\label{eqCohVect},
\end{equation}
with ${\vec{C}_\mathrm{F}}_n^l$ the \refere{4-element vector of the flux in each beam}, defined for each aperture $i$ by:
\begin{equation}
 {C_\mathrm{F}}_n^{l \; (i)}= \left|A_n^{l \; (i)}\right|^2 ,
\end{equation}
and ${\vec{C}_\mathrm{C}}_n^l$ the \refere{6-element} complex coherence vector, defined for each baseline $k$ combining telescopes $(i,j)$ by:
\begin{equation}
 {C_\mathrm{C}}_n^{l \; (k)}= A_n^{l \; (i)} {A_n^{l \; (j)}}^* .
\end{equation}

\begin{table}
  \caption{Simulated phase shifts of the B-A quadrature (nominal value of 90~\degr) for the 6 baselines. The first line presents the average of the phase shifts over the spectral channels. The second line presents the variation range over the spectral channels.} 
 \label{ABCD} 
 \centering 
 \begin{tabular}{c c c c c c c c } 
  \hline\hline 
  \multicolumn{2}{c}{Baseline} 		&1-2&1-3&1-4&2-3&2-4&3-4\\
  \hline
  Phase & ($\lambda_0$)  		&92 &94 &95 & 103& 107& 79\\
  shifts ($^\circ$)& ($\Delta\lambda$) 	&2  &15  &15  & 7 & 9& 11\\
  \hline 
 \end{tabular}
\end{table}

Finally, two sources of noise are added to each pixel of the image $\tens{I}_n$: 
\begin{itemize}
 \item Photon noise, amplified by a factor $F_{APD}=1.5$ to take into account the noise excess factor from an avalanche photo-diode detector (APD);
 \item Read-out noise of $\mathrm{RON}=4~\mathrm{e}^{-}$ rms, amplified by a factor $\sqrt{N_{pix}}$ to account for a spread of the intensities over $N_{pix}=2$ detector pixels due to non-perfect imaging optics.
\end{itemize}
These parameters are the characteristics of the SELEX-Galileo detector of the GRAVITY fringe tracker \citep{Finger2010}.

\section{Description of the fringe tracker algorithms\label{algo}}

Fringe tracking \refere{algorithms} can be split into two equally important parts:
\refere{the phase sensor, whose role is to provide accurate estimations of the OPD on all baselines, and the controller, which compute commands to the piston actuators out of these estimates. We provide in this section a comprehensive description of the algorithms used in our simulations.}

\subsection{OPD estimation \label{opd_est}}

The \refere{6-baseline} residual OPD vector $\bhdelta_n$ at time step $n$ is computed from the last available image $\tens{I}_{n-1}$ (cf. Fig.~\ref{Time_diag}). Two operators are used for its estimation. 

\refere{The phase delay operator (PD) estimates the OPD from the phase of the wide-band intensities. It provides an instantaneous estimation of the OPD, but modulo the wide-band wavelength (2.2~$\mu$m in the K band), and is thus alone insensitive to larger OPD fluctuations. To estimate the absolute OPD on each baseline, we thus use the group delay estimator (GD), which localizes the OPD of the maximum of the coherence envelope from the dispersed intensities.}

\subsubsection{Phase delay estimation \label{PD_est}}
\refere{For the PD estimation, we first compute the 24-element} vector $\vec{I}_{n-1}^{\mathrm{wb}}$ \refere{of the wide-band intensities from the dispersed image}:
\begin{equation}
 \vec{I}_{n-1}^{\mathrm{wb}}=\sum_{l=1}^{N_\lambda}\vec{I}_{n-1}^l .
\end{equation}
\refere{Assuming that the beam combiner has been preliminarily characterized and that the wide band and dispersed V2PMs have been perfectly measured, the 16-element} vector $\vec{\hat{C}}_n^{\mathrm{wb}}$ \refere{of the wide-band coherences} is then estimated from the pixel-to-visibility matrix (P2VM) $\tens{P2VM}^{\mathrm{wb}}$, pseudo-inverse matrix of the \refere{corresponding} V2PM:
\begin{equation}
 \vec{\hat{C}}_n^{\mathrm{wb}}=\tens{P2VM}^{\mathrm{wb}} \, \vec{I}_{n-1}^{\mathrm{wb}}.
\end{equation}
The residual PD OPD vector is deduced from the phase of \refere{the complex coherences ${\vec{\hat{C}}_{\mathrm{C}\,n}}^{\mathrm{wb}}$, i.e the 12 last elements of the wide-band coherence vector (see Eq.~\ref{eqCohVect} for the structure of the coherence vector)}:
\begin{equation}
 \bhdelta_n^{\mathrm{PD}}=\frac{\lambda_0}{2\pi}\arg\left({\vec{\hat{C}}_{\mathrm{C}\,n}}^{\mathrm{wb}}\right) \pmod{\lambda_0}.
\end{equation}

The resulting PD OPDs are thus computed modulo \refere{the effective wide-band wavelength $\lambda_0=2.2~\mu$m in our simulations.}

\subsubsection{Group delay estimation\label{GD_est}}
To estimate the position of the envelope of coherence with the GD estimator, the fringes are spectrally dispersed over $N_\lambda$ spectral channels, and the group delay OPDs are estimated from the coherence measurements on these spectral channels. \refere{As the signal-to-noise ratio (S/N) of the synthetic wide-band intensities is increased by summing the $N_\lambda$ spectral channels, we start by adding the last $N_\lambda$ images available at iteration $n$ to obtain an image $\tens{\tilde{I}}_{n-1}$ with an increased S/N}: 
\begin{equation}
 \tens{\tilde{I}}_{n-1}=\sum_{n'=1}^{N_\lambda}\tens{I}_{n-n'}.\label{eqSum}
\end{equation}
\refere{The consequence of this operation is that the GD estimator do not measure the OPD at frame $n-1$ but its average over the last five iterations.}

\refere{To estimate the GD OPDs, we use} an algorithm similar to the double Fourier interferometry used at IOTA \citep{Pedretti2005}. 
The \refere{16-element} coherence vector $\vec{\hat{C}}_n^{l}$ is estimated at time step $n$ for each spectral channel $l$ using \refere{the calibrated} $\tens{P2VM}^{l}$: 
\begin{equation}
 \vec{\hat{C}}_n^{l}=\tens{P2VM}^{l} \, \vec{\tilde{I}}_{n-1}^{l},
\end{equation}  
\refere{then the corresponding 6-element complex coherent vector $\vec{\hat{C}}_{\mathrm{C}\,n}^{l}$ is extracted from its 12 last elements.}

The cross-spectrum products between the complex coherences of each adjacent spectral channel are computed. The cross-spectrum product $\vec{X}_n(l_1,l_2)$ between channels $l_1$ and $l_2$ is defined by the \refere{element-wise} product of the first one with the conjugate of the second one:
\begin{equation}
 \vec{X}_n(l_1,l_2)={\vec{\hat{C}}_{\mathrm{C}\,n}}^{l_1} \cdot \left( {\vec{\hat{C}}_{\mathrm{C}\,n}}^{l_2} \right)^*.
\label{xp}
\end{equation}
\refere{Estimations of the group delay OPDs can be directly computed from the phase of each cross-spectrum product between adjacent spectral channels, within a range of $\Lambda_{(l,l+1)}$:}
\begin{equation}
\bhdelta_n^{\mathrm{GD}}(l,l+1)=\frac{\Lambda_{(l,l+1)}}{2\pi}\arg \left(\vec{X}_n(l,l+1) \right) \pmod{\Lambda_{(l,l+1)}},
\end{equation}
where $\Lambda_{(l,l+1)}$ is the beating wavelength between the spectral channels $l$ and $l+1$:
\begin{equation}
 \Lambda_{(l,l+1)}=\frac{\lambda_l\lambda_{l+1}}{\lambda_{l+1}-\lambda_l}.
\end{equation}
To increase the precision, \refere{we actually compute} the GD OPD vector from the average of these $N_\lambda-1$ \refere{individual GD OPD vectors}:
\begin{equation}
 \bhdelta_n^{\mathrm{GD}}=\frac{1}{N_\lambda-1}\sum_{l=1}^{N_\lambda-1}\bhdelta_n^{\mathrm{GD}}(l,l+1).
\end{equation}

\refere{This estimator becomes inaccurate as soon as the OPD is larger than $\pm \min(\Lambda_{(l,l+1)})/2$ because of the average of OPDs wrapped at different values. The fringe tracker needs to be designed with a spectral resolution enough to avoid that such a problem happens within the range of variation of the OPD disturbance. For the fringe tracker of GRAVITY, the spectral resolution of $R=22$ gives a range of validity of $\pm 16~\mu$m for the group delay estimator, which is largely enough to measure for the OPD disturbance, all the more so as when the loop is closed, even if the tracking is lost during several iterations. In addition, with this large range of validity, the GD OPD estimator can also be used to search for the fringes before closing the control loop, by moving the actuators over a long stroke then start the control with the actuators at positions close to the null OPDs.}

\subsubsection{Final OPD estimation}
To solve for the ambiguity \refere{on the position of the fringe tracked by the PD estimator, we compute the final} OPD $\hat{\delta}_n^{(k)}$ on each baseline $k$ \refere{as}:
\begin{equation}
 \hat{\delta}_n^{(k)}=\left \{ \begin{array}{rl}
                                  \hat{\delta}_n^{\mathrm{PD}~(k)} & \mathrm{if}\quad \left| \hat{\delta}_n^{\mathrm{GD}~(k)} \right| < \lambda_0/2 \\
				  \hat{\delta}_n^{\mathrm{GD}~(k)} & \mathrm{otherwise}
                                 \end{array}
			  \right. . \label{final_opd}
\end{equation}
The \refere{zero-}phase of the fringe is thus tracked except when a fringe jump occur, \refere{in which case the group delay is used to bring the fringes back to the maximum of coherence}. The system is therefore stabilized to the zero-phase of the central fringe, enabling long integrations on \refere{a dedicated separate science detector} with no loss in the visibility accuracy \refere{and thus observation of faint targets}.

\subsection{OPD uncertainty estimation \label{noise_est}}

\refere{For interferometer combining more than three telescopes, the redundancy between the closure phase relations \citep{Jennison1958} provides a simple way to verify that the estimated piston on each aperture are compatible altogether when observing a unresolved target. If they are not, this redundancy can be used to improve their estimation, by identifying the OPDs with the greatest uncertainty. The controllers implemented in our simulations and described in Sec.~\ref{controllers} make use of this particularity. In this purpose, the uncertainties on the OPDs also need to be estimated at each loop iteration.}

\refere{For this, we first estimate} the variance $\tens{\hat{\sigma}_{I\,n-1}}^2$ \refere{on the intensity} of each pixel of the last available image $\tens{I}_{n-1}$:
\begin{equation}
 \tens{\hat{\sigma}_{I\,n-1}}^2=F_{APD} \tens{I}_{n-1} + N_{pix} \mathrm{RON}^2 ,
\end{equation}
\refere{expression which takes} into account the detector characteristics as described in Sect.~\ref{BC}. The first part of the equation corresponds to the photon noise amplified by the noise excess factor of the detector, and the second part corresponds to the detector read-out noise on $N_{pix}$ pixels.

\refere{This variance map is then used to estimate the uncertainty for both PD and GD estimators.}

\subsubsection{PD OPD \refere{uncertainty}}
\refere{For the uncertainties on the PD estimations, we compute} the \refere{$24\times24$} covariance matrix $\tens{\Sigma_I}_{n-1}^{\mathrm{wb}}$ of the wide-band intensities, \refere{assuming that the intensities are uncorrelated}:
\begin{equation}
 \tens{\Sigma_{I\,n-1}}^{\mathrm{wb}}= \mathrm{diag}\left(\sum_{l=1}^{N_\lambda}\left(\vec{\hat{\sigma}_{\tens{I}\,n-1}}^{l}\right)^2 \right).
\end{equation}
We then deduce the covariance matrix of the wide-band complex coherence:
\begin{equation}
 \tens{\Sigma_{C\,n}}^{\mathrm{wb}}=\tens{P2VM}^{\mathrm{wb}} \, \tens{\Sigma_{I\,n-1}}^{\mathrm{wb}} \, \left(\tens{P2VM}^{\mathrm{wb}}\right)^{\rm H},
\end{equation}
where $\tens{M}^{\rm H}$ is the adjoint matrix of $\tens{M}$. \refere{We select} the variance $\left({{\hat{\sigma}_{C\,n}}^{\mathrm{wb} \; (k)}}\right)^2$ of the complex coherence of the baseline $k$ as the corresponding diagonal term of the covariance matrix:
\begin{equation}
 \left({{\hat{\sigma}_{C\,n}}^{\mathrm{wb} \; (k)}}\right)^2 = \tens{\Sigma_{C\,n}}^{\mathrm{wb} \; (k,k)}.
\end{equation}

The PD OPD \refere{uncertainty} vector ${\bhsigma_{PD}}_n$ is finally estimated from the \refere{uncertainty} $\bhsigma_\mathrm{\phi\; n}^\mathrm{wb}$ on the phase of the complex coherence (see Appendix~\ref{phi_noise_app} for the derivation of the noise on the phase of the complex coherence):
\begin{equation}
 {\bhsigma_{PD}}_n = \frac{\lambda_0}{2\pi}\bhsigma_\mathrm{\phi\; n}^\mathrm{wb} .
\end{equation}

\subsubsection{GD OPD \refere{uncertainty}}
\refere{Similarly,} the covariance matrix $\tens{\Sigma_I}_{n-1}^{l}$ of the intensities of spectral channel $l$ is computed from the sum of the $N_\mathrm{\lambda}$ last pixel noise estimations:
\begin{equation}
 \tens{\Sigma_{I\,n-1}}^{l}= \mathrm{diag}\left(\sum_{n'=1}^{N_\mathrm{\lambda}}\left(\vec{\hat{\sigma}_{\tens{I}\,n-n'}}^{l}\right)^2 \right).
\end{equation}
We then deduce the covariance matrix of the complex coherence of each spectral channel $l$:
\begin{equation}
 \tens{\Sigma_{C\,n}}^{l}=\tens{P2VM}^{l} \, \tens{\Sigma_{I\,n-1}}^{l} \, \left(\tens{P2VM}^{l}\right)^{\rm H}.
\end{equation}
The vector ${{\bhsigma_{\mathrm{C}\,n}}^l}^2$ of the variance of the complex coherences is computed as the diagonal terms of this covariance matrix:
\begin{equation}
 \left({{\hat{\sigma}_{\mathrm{C}\,n}}^{l \; (k)}}\right)^2 = \tens{\Sigma_C}_n^{l \; (k,k)}.
\end{equation}
The variance vector ${{\bhsigma_\mathrm{X}}_n}^2(l,l+1)$ of the cross-spectrum product between each adjacent spectral channels $l$ and $l+1$ is derived as described in Appendix~\ref{Noise_Xp}. 

The GD OPD variance vector is then computed from the average of the $N_\lambda-1$ variances ${\bhsigma_\mathrm{\phi\; n}^{l,l+1}}^2$ of the phase of the cross-product between each adjacent spectral channels (see Appendix~\ref{phi_noise_app} for the derivation of the noise on the phase of the cross product):
\begin{equation}
 {\bhsigma_{GD}}_n = \frac{1}{N_\lambda-1} \sqrt{\sum_{l=1}^{N_\lambda-1}\left(\frac{\Lambda_{(l,l+1)}}{2\pi}\right)^2 {\bhsigma_\mathrm{\phi\; n}^{l,l+1}}^2}.
\end{equation}

\subsubsection{OPD \refere{uncertainty}}
\refere{Finally,} the \refere{uncertainty on the estimation of the OPD} on each baseline $k$ is determined \refere{by selecting the uncertainty of the corresponding estimator,} in the same way as \refere{we select the final OPD in Eq.~\ref{final_opd}}:
\begin{equation}
 \hat{\sigma}_n^{(k)}=\left \{ \begin{array}{rl}
                                  {\hat{\sigma}_{PD\,n}}^{(k)} & \mathrm{if}\quad \left| \hat{\delta}_n^{\mathrm{GD}~(k)} \right| < \lambda_0/2 \\
				  {\hat{\sigma}_{GD\,n}}^{(k)} & \mathrm{otherwise}
                                 \end{array}
			  \right. .
\end{equation}

\subsection{OPD controllers\label{controllers}}
\refere{Once the OPDs and their uncertainties have been estimated on each baseline, they can be used and combined by the controller to compute optimal commands to be applied to the piston actuators and correct for the residual errors.} In this section we describe the different controllers compared in our simulations: an integrator controller and a controller based on the Kalman algorithm.

\subsubsection{Integrator controller \label{secIntegrator}}
The integrator controller is implemented in two different schemes to investigate the better way to compute four piston commands from six estimated OPDs. In the first scheme (hereafter called \emph{OPD control scheme}), commands are computed in baseline space to correct for the residual OPDs, then \refere{reverted to piston commands.} 
In the second scheme (the \emph{piston control scheme}), residual pistons are estimated from the estimated OPDs, then actuator commands are computed to correct for the piston residuals.

\refere{To account for the latency of the GD estimation resulting from the sum of $N_\lambda$ previous images (Eq. \ref{eqSum})}, two different integrator  gains are used, $K_\mathrm{PD}$ and $K_\mathrm{GD}$. The optimum gain on the baseline $(k)$ is set such that:
\begin{equation}
 K_n^{(k)} = \left \{ \begin{array}{rl}
			K_\mathrm{PD} & \mathrm{if}\quad \left| \hat{\delta}_n^{\mathrm{GD}~(k)} \right| < \lambda_0/2\\
			K_\mathrm{GD} & \mathrm{otherwise}
		      \end{array}\right. ,
\end{equation}
\refere{defining} $\vec{K}_n$ as the vector of gains on the six baseline. $K_\mathrm{PD}$ and $K_\mathrm{GD}$ were \refere{preliminarily} optimized with closed-loop simulations on a grid of values for both gains. \refere{We determined the optimal gain combination by minimizing the sum over all the baselines and the whole temporal sequence of the squared residual OPDs.}

Moreover, to take advantage of the redundancy in the 4 telescopes -- 6 baselines architecture, weighted recombinations of the OPD residuals are computed \refere{from the estimated OPD uncertainties on} the 6 baselines \citep{Menu2012}. \refere{If $\tens{M}$ is the $6\times4$} transfer matrix computing the 6-element OPD vector from a 4-element piston vector $\vec{\hat{p}}_n$:
\begin{equation}
 \bhdelta_n = \tens{M} \, \vec{\hat{p}}_n,
\end{equation}
the weighted recombination of the OPDs is achieved by computing the weighted generalized inverse $\tens{M}_n^{\mathrm{W}~\dagger}$ of the matrix M:
\begin{equation}
 \tens{M}_n^{\mathrm{W}~\dagger} = \left(\tens{M}^{\rm T} \, \tens{W}_n \, \tens{M}\right)^\dagger \, \tens{M}^{\rm T} \, \tens{W}_n,
\end{equation}
with the weight matrix $\tens{W}_n$ defined as the diagonal matrix of the \refere{inverse of the OPD variance vector} $\bhsigma_n^2$:
\begin{equation}
 \tens{W}_n=\mathrm{diag}\left( 1 / \bhsigma_n^2 \right).
\end{equation}
This weighted combination is used in both the OPD and piston scheme of the integrator controller.

\paragraph{OPD control scheme:}~\\
In this configuration, the weighted recombination of the OPDs $\bhdelta_n^W$ is first computed:
\begin{equation}
 \bhdelta_n^W = \tens{1}_n^\mathrm{W} \, \bhdelta_n , \label{weight_rec}
\end{equation}
with the weighted identity matrix
\begin{equation}
 \tens{1}_n^\mathrm{W} = \tens{M} \, \tens{M}_n^{\mathrm{W}~\dag} \label{Widentity}.
\end{equation}

OPD commands $\vec{u}_n^\mathrm{opd}$ are then computed to compensate for these residual OPDs:
\begin{equation}
 \vec{u}_n^\mathrm{opd}=\vec{K}_n  \, \bhdelta_n.
\end{equation}
They are lastly inverted to compute the piston commands $\vec{U}_n$:
\begin{equation}
 \vec{U}_n=\vec{U}_{n-1}+\tens{M}_n^{\mathrm{W}~\dag} \, \vec{u}_n^\mathrm{opd}.\label{EQposi1}
\end{equation}

\paragraph{Piston control scheme:}~\\
In the aperture space control configuration, the residual piston vector $\vec{\hat{p}}_n$ is first deduced with the weighted inversion of the OPD vector $\bhdelta_n$:
\begin{equation}
 \vec{\hat{p}}_n=\tens{M}_n^{\mathrm{W}~\dag} \, \bhdelta_n.
\end{equation}

The correction $\vec{U}_n$ of the integrator controller is then computed and sent to the piston actuators:
\begin{equation}
 \vec{U}_n=\vec{U}_{n-1}+\left(\tens{N} \, \vec{K}_n \right) \, \vec{\hat{p}}_n.\label{EQposi2}
\end{equation}
The matrix $\tens{N}$ converts the baseline gain vector $\vec{K}_n$ into an aperture gain vector by averaging the gain on the three baselines related to each aperture:
\begin{equation}
 \tens{N}= \frac{1}{3} \left|\tens{M}^{\rm T}\right|.
\end{equation}

\subsubsection{Kalman controller \label{Kalman_controller}}

\refere{The Kalman filter is a recursive algorithm which can predict the new state of the disturbance based on a model of their evolution, which have to be identified beforehand, and from a set of previous measurements that are used to compare the prediction and the real measurements. From the estimation of the statistical characteristics of the disturbance, their evolution model, and the uncertainty on the measured residuals, control commands can be computed in a statistically optimal way.}

\refere{The Kalman controller used in our simulations is very similar to the algorithm described in \citet{Menu2012}, except from a few adaptations to our comprehensive case. We remind here the principle of the Kalman filter and give the details in which its implementation differs from this reference.}

\paragraph{State-space formalism:}~\\
\refere{The formalism used by a Kalman filter is based on a couple of assumptions: the disturbance  can be described by a set of independent components (typically the atmospheric piston and a discrete number of longitudinal vibrations for fringe tracking systems) and the evolution each component can be described by an iterative linear model.}

\refere{Here we follow \citet{Meimon2010} statement that the evolution of both the atmospheric and the vibration components can be describe by an autoregressive model of order two of the form:
\begin{equation}
\delta_{n+1}^v= a_1^v \delta_{n}^v +  a_2^v \delta_{n-1}^v + v_n^i,
\end{equation}
where $\delta_n^v$ is the $v$-th component of the disturbance at time step $n$, the coefficients $a_1^v$ and $a_2^v$ are computed from the component characteristics (natural frequency and damping coefficient, the later being lower than 1 for vibrations and greater than 1 for the atmospheric turbulence) and $v_n$ a white noise of standard deviation $\sigma^v$ triggering the excitation of the component \citep[see][for the derivation of the coefficients $a_1^v$ and $a_2^v$]{Meimon2010}. In our simulated case, given $N_{comp}$ the total number of disturbance components, all baselines included, this evolution model can be generalized to the following matrix form:
\begin{equation}
\vec{x}_{n+1}=\tens{A} \vec{x}_{n}+ \vec{v}_n,
\end{equation}
where $\vec{x}_{n}$ is a $2 N_{comp}$-element vector describing the state of each component at iteration $n$ and $n-1$, A is the $2 N_{comp} \times 2 N_{comp}$ matrix gathering the $a_1$ and $a_2$ coefficients of every component, and $\vec{v}_n$ is the $2 N_{comp}$-element vector of the excitation noises.}

\refere{The OPD vector measured at iteration $n$ is related to the state vector by the $6 \times 2 N_{comp}$ matrix $\tens{C}$ (which sums the $\delta_{n-1}^v$ components related to each baselines in the state vector $\vec{x}_{n}$), the OPDs induced by the actuator positions, and the measurement noise at iteration $n$:
\begin{equation}
\bdelta_n = \tens{C} \vec{x}_n - \tens{M} \vec{U}_{n-2} + \bsigma_n .
\end{equation}}

\paragraph{Control algorithm:}~\\
\refere{Assuming that the evolution model of the OPD disturbance has been preliminarily identified (i.e. the matrices $\tens{A}$, $\tens{C}$ and the excitation noises), the Kalman controller algorithm consists of four successive steps:
\begin{enumerate}
\item We compute the error $\vec{e}_n$ between the  OPDs $\bhdelta_n^W$ estimated at iteration $n$ and the OPDs predicted at the previous iteration by the Kalman controller. The latter result from the effective actuator positions delayed by two iterations (See Fig.~\ref{Time_diag}) and from the disturbance state $\vec{\hat{x}}_{n|n-1}$ predicted by the controller for iteration $n$ based on all observations up to iteration $n-1$:
\begin{equation}
\vec{e}_n = \bhdelta_n^W - \left( \tens{C} \vec{\hat{x}}_{n|n-1} - \vec{U}_{n-2} \right).
\end{equation}
\item The state vector is then updated with this correction vector and the Kalman gain matrix $\tens{G}$:
\begin{equation}
\vec{\hat{x}}_{n|n}=\vec{\hat{x}}_{n|n-1}+\tens{G}  \vec{e}_n.
\end{equation}
\item The next state of the disturbance is predicted using the evolution model defined by matrix $\tens{A}$:
\begin{equation}
\vec{\hat{x}}_{n+1|n}=\tens{A} \vec{\hat{x}}_{n|n}.
\end{equation}
\item The command vector $\vec{U}_n$ to the actuators is finally computed by inverting the optimal OPD commands to optimal piston commands, with respectively matrices $\tens{K}$ and $\tens{M}_n^{\mathrm{W}~\dag}$:
\begin{equation}
\vec{U}_n =\tens{M}_n^{\mathrm{W}~\dag}  \tens{K}  \vec{\hat{x}}_{n+1|n}\label{EQposi3},
\end{equation}
with $\tens{K}$ a $6 \times 2 N_{comp}$ matrix which sums the $\delta_{n+1}^v$ components related to each baselines in the state vector $\vec{\hat{x}}_{n+1|n}$.
\end{enumerate}}

\refere{The Kalman gain $\tens{G}$ is a $2 N_{comp} \times 6$ weight matrix that determines how much confidence in the theoretical model we have with respect to real measurements. It is statistically optimal when computed as follows, knowing the OPD uncertainties and the disturbance excitations:
\begin{equation}
\tens{G} = \tens{\Sigma}_\infty \tens{C}^{\rm T} \left( \tens{C} \tens{\Sigma}_\infty \tens{C}^{\rm T} + \tens{\Sigma}_w \right)^{-1},\label{eqGainKal}
\end{equation}
with $\tens{\Sigma}_\infty$ the solution of the Riccati equation:
\begin{equation}
\tens{\Sigma}_\infty = \tens{A} \tens{\Sigma}_\infty \tens{A}^{\rm T} - \tens{A} \tens{\Sigma}_\infty \tens{C}^{\rm T} \left( \tens{C} \tens{\Sigma}_\infty \tens{C}^{\rm T} + \tens{\Sigma}_w \right)^{-1} \tens{C} \tens{\Sigma}_\infty \tens{A}^{\rm T} +\tens{\Sigma}_v.
\end{equation}
$\tens{\Sigma}_v$ is the $2 N_{comp} \times 2 N_{comp}$ covariance matrix of the excitation noise of the disturbance components, simply a diagonal matrix in our system, with the $\sigma^v$ terms at the $a_1^v$ positions and zero terms at the $a_2^v$ positions, and $\tens{\Sigma_w}$ is the covariance matrix of the OPD measurement noise. Since the OPDs are estimated from two different estimators (Eq.~\ref{final_opd}), we used two different measurement noise covariance matrices $\tens{\Sigma}_w$, computed with a similar weighted combination as for the OPDs. We thus used two different Kalman gain matrices, optimized respectively for phase and group delay OPD measurements, the proper gain for each baseline being selected the same way as in Eq.~\ref{final_opd}.} 

\paragraph{Model identification:}~\\
\refere{To identify the evolution model of the disturbance, a sequence of OPDs representative from the variations has to be measured before tracking the fringes with the Kalman controller. To do so, we used a similar approach than those described in \citet{Menu2012} and \citet{Meimon2010}, with a few adaptation to our case.}

\refere{Because of the large OPDs induced but the atmospheric disturbance, open-loop measurements can not be obtained without losing the fringes. Instead, a pseudo-open loop (POL) OPD sequence has to be computed, by tracking the fringes with a classical controller to measure the OPD residuals with high precision, then reconstructing the corresponding disturbance sequence knowing the actuator positions. In our simulations, we computed POL sequences by performing short fringe tracking simulations with the piston scheme integrator controller described in Sec.~\ref{secIntegrator}. To compute the POL OPD sequence, we used the OPDs estimated in Eq.~\ref{final_opd}, actuator commands computed in Eq.~\ref{EQposi2}, and used the weighted identity matrix computed in Eq.~\ref{Widentity} to have estimates of the POL OPDs improved by the redundant configuration of instrument:}
\begin{equation}
\bdelta_n^{POL} = \tens{1}_n^\mathrm{W}  \times (\bhdelta_n + \tens{M} \times \vec{U}_{n-2}).
\end{equation}

\refere{From these POL sequences, we used the same method as in \citet{Menu2012} to identify the disturbance components and their evolution model, and compute the matrices $\tens{A}$, $\tens{\Sigma}_v$, $\tens{C}$, and $\tens{K}$. The measurement covariance matrices $\tens{\Sigma}_w$ for the phase delay and group delay estimators also are computed from the POL OPD sequences.}

\section{Results of the simulations\label{results}}

The performances of the integral and of the Kalman controllers were simulated for different star magnitudes, observing conditions, and loop frequencies. The \refere{varying} parameters are summarized in Table~\ref{var param}. Both versions (piston and OPD scheme) of the integral controller were simulated, and the disturbance model used for the Kalman controller was identified with POL sequences of four different lengths.

To improve the statistics on the results, each simulation is performed 10 times with different \refere{random} disturbance sequences \refere{for each baseline}\footnote{\refere{The typical computation time for one realization is about 1h for simulation with every controller (two integrator and Kalman controller with four different POL-sequence lengths) at a given star magnitude and loop frequency on a 8-core 2.27~GHz Intel Xeon machine. Our simulator is a high-level research tool intended to test fringe tracking algorithms, and has not been optimized for computation celerity.}}. \refere{The sequences of true OPD residual are computed for each baseline and each realization as the difference between the simulated disturbance sequence and the sequence of OPDs applied by the piston actuators. These positions correspond to the vector $\vec{U}_n$ computed in Eqs.~\ref{EQposi1}, \ref{EQposi2}, and \ref{EQposi3} with one iteration of delay (see Fig.~\ref{Time_diag}). The standard deviations of these sequences\footnote{\refere{The first 1000 iterations are excluded from the standard deviation computation to account for the time needed by the controller to converge.}} over time are then computed and the performance of the controllers is judged with the median of these standard deviations over the 6 baselines and the 10 random realizations (thus median over 60 observables). } 

In the following, we present results at the loop frequency minimizing these standard deviation of the OPD residuals. The \refere{optimal frequencies are also presented for each star magnitude, each controller and observing conditions}, and provide the best compromise between the controller bandwidth and accuracy of the OPD estimation \refere{depending of the reference star magnitude}.

\subsection{Performances at different vibration levels}
\begin{figure}
 \resizebox{\hsize}{!}{\includegraphics{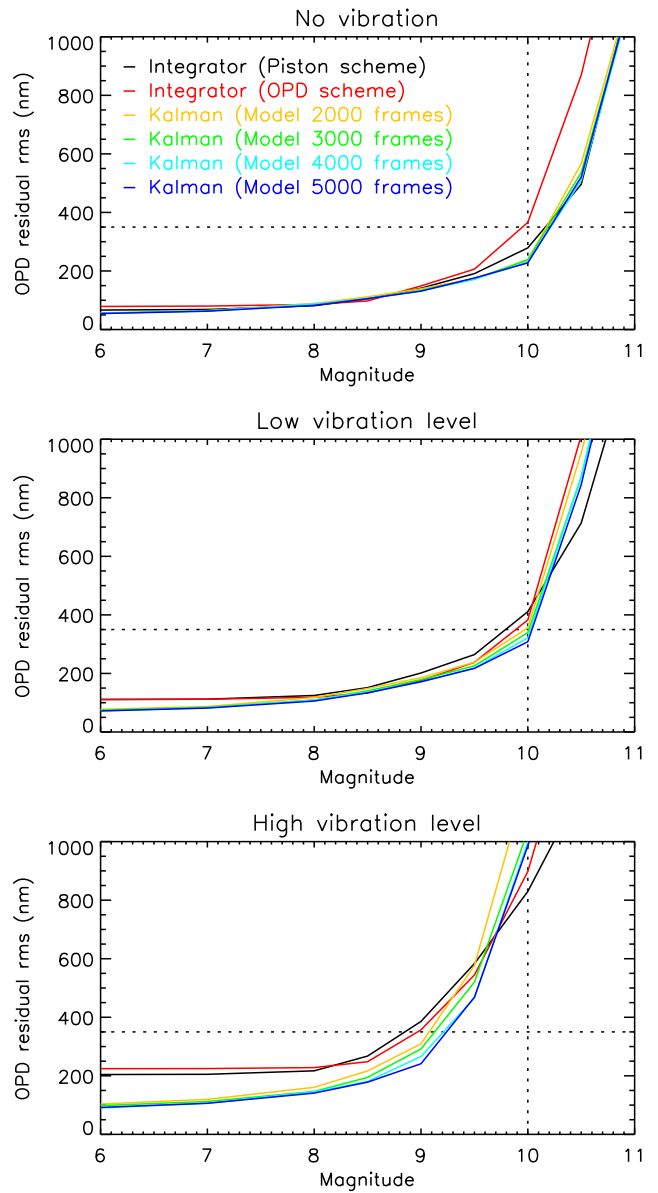}}
 \caption{OPD residuals as a function of magnitude \refere{for three different vibration levels}. \tmp{Top}: no vibration; middle: \refere{low vibration level with 150~nm rms OPD}; \tmp{bottom}: \refere{high vibration level from 240 to 380~nm rms OPD}. The atmospheric \refere{turbulence} is 10~$\mu$m rms OPD and, and the residual tip-tilt of the beams is 15~mas rms. \refere{The dashed lines correspond to the specification of the fringe tracker of GRAVITY, which is to stabilize the fringes down to 350~nm rms on a reference star of magnitude $K=10$.}}
 \label{simu_vib_levels}
\end{figure}

\begin{figure}
\resizebox{\hsize}{!}{\includegraphics{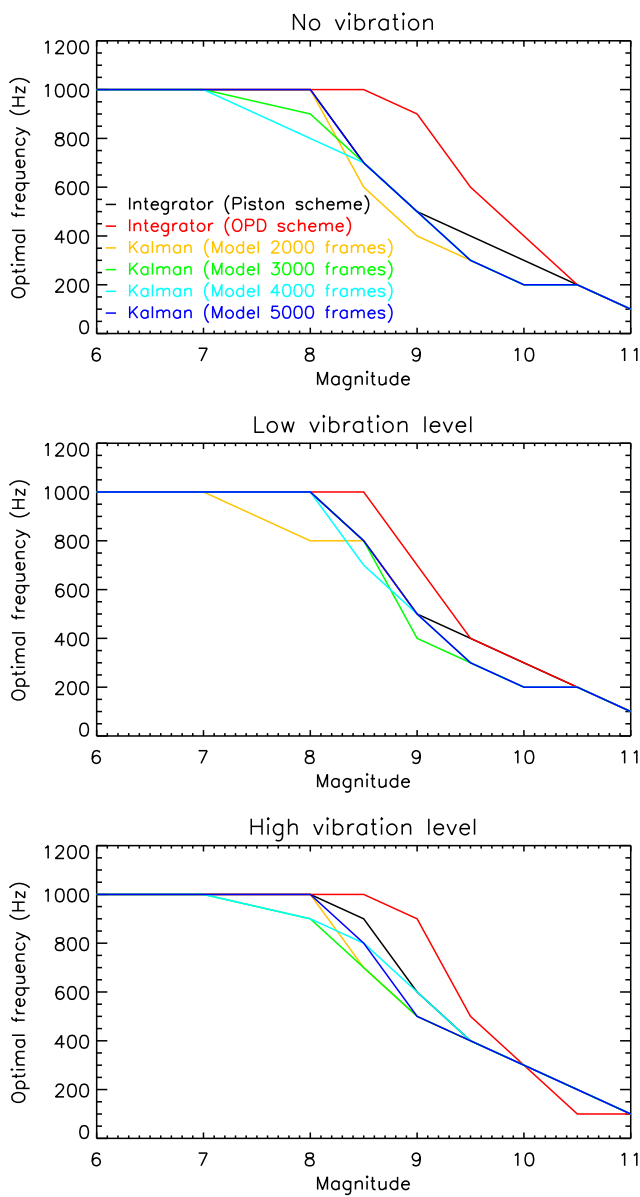}}
 \caption{Optimal frequencies as a function of the magnitude for conditions related to Fig.~\ref{simu_vib_levels}.}
 \label{simu_vib_levels_freq}
\end{figure}

We analyzed the performances of the controllers by simulating three different vibration levels: no longitudinal vibration, a low vibration level \refere{corresponding to the one expected at VLTI in late} 2014 (150~nm rms OPD on each baseline), and \refere{a high} vibration level \refere{similar to the one} currently estimated at VLTI on the UTs (cf. Tables~\ref{vibs} and \ref{vibs_rms}). The residual OPDs at optimal loop frequency are presented in Fig.~\ref{simu_vib_levels} as a function of magnitude for each vibration levels \refere{and} the corresponding optimal frequencies are presented in Fig.~\ref{simu_vib_levels_freq}. \refere{Flux variations resulting from 15~mas rms residual tip-tilt and atmospheric turbulence with $10~\mu$m rms OPD were simulated for each case}.

For reference stars \refere{brighter than} $K\sim$9.5, the piston and OPD schemes of the integrator are equally efficient, for the three vibration levels. The piston scheme integrator is more robust to fainter magnitudes and presents a gain of a few hundred of nanometers rms over the OPD scheme.

Without vibrations (Fig.~\ref{simu_vib_levels}, top), the Kalman controller and the piston scheme integrator have very similar performances, whatever the magnitude. \refere{For the simulated flux variation level, both controller are thus equally efficient to compensate for the atmospheric disturbance. In addition, with $10~\mu$m rms of atmospheric OPD, a 2000-frame POL sequence is as efficient as longer ones, and is then enough to properly identify the turbulence component in the disturbance even at low S/N.}

At high S/N, the Kalman controller is mainly insensitive to the vibration level, unlike the integrator controllers. 
\refere{For stars brighter than $K\sim8$, the OPD residuals with the Kalman controller are stable to $100 \pm50$~nm rms whatever the vibration level, whereas the performance of the integrator controller clearly scales with the vibration level. The Kalman controller is thus very efficient to cancel out the vibrations in this regime. In addition, at high S/N, a 2000-frame POL sequence is enough to properly identify and calibrate the vibrations that we simulated. If the system present vibrations with lower frequencies, longer POL sequences might however be needed to properly characterize their evolution model. Performances of all controller are limited by the maximal simulated loop frequency of 1~kHz (see Fig.~\ref{simu_vib_levels_freq}). Using a higher loop frequency in this case may actually improve the fringe stabilization by increasing the bandwidth of the controller.}

\refere{At medium S/N regime, the performances of all controllers continuously decrease, because the estimated OPDs are less accurate and the error on the piston commands are then greater. The Kalman controller still stabilizes the fringes better than the integrator thanks to its predictive algorithm which optimally weights predictions from the disturbance model and measured OPD residuals. However, it become slightly less efficient in this regime than at high S/N because the state vector is updated with measurements of poor precision, and because the disturbance model is identified from POL sequences where the fringes were stabilized with a classical integrator, which is poorly efficient in the mid-S/N regime. The disturbance sequences reconstructed from the estimated OPD residuals are not as accurate as when the fringes are efficiently stabilized, and the disturbance components are not properly characterized. This is particularly observable in presence of strong vibrations, for two principal reasons. First, a high level of vibrations significantly increase the total OPD fluctuations and leads to a poor fringe tracking during the POL sequence, and thus to an even less accurate disturbance model. Secondly, the sharper the vibration, the more precise its identification must be to provide an accurate control. For very narrow-band components ($k\ll1$), a slight error in its characterization (e.g its natural frequency) can lead to a poor compensation.  On the contrary, the atmospheric disturbance is properly corrected by the Kalman controller even at very low S/N (see Fig.~\ref{simu_vib_levels}, top), because its model with a large damping coefficient $k$ makes its identification very robust to large uncertainties.}

At low S/N, the Kalman controller becomes less robust than the integrator controller if there are vibrations in the system. \refere{With OPD residuals greater than 500~nm rms with the integrator controller, the reconstructed POL sequences are not accurate enough to provide a proper model identification for the vibrations, and the Kalman controller can not converge to correct commands. In addition, at low S/N the controller loop frequency lower than 400~Hz for magnitudes above 9.5 makes identification of vibrations with high frequency more difficult.}

The performances of the Kalman controller slightly improve if the length of the POL sequence is increased. The longer the POL sequence, the better the precision in the identification of the disturbance components. \refere{However, using longer POL sequences may reveal to be useless in practice: the disturbances may possibly vary during the observation, and short POL sequences are less subject to model variations. In our simulations, we assume temporally invariant disturbances, but we discuss this possibility in Sec.~\ref{discussion}.}

\subsection{Performances at different flux variation levels}

\refere{When working at the sensitivity limit of a fringe tracker, flux variations are a serious cause of performance loss with classical controllers: if the flux in one beam drops too low to accurately estimate the OPDs with a sufficient precision (called a flux dropout hereafter), incorrect or no commands at all are sent to the actuator. The typical sources of flux variations for single-mode fibered interferometer are the residual wavefront errors from the AO systems, telescope tip-tilt vibrations, and tip-tilt errors from the guiding system.}

\begin{figure}
 \resizebox{\hsize}{!}{\includegraphics{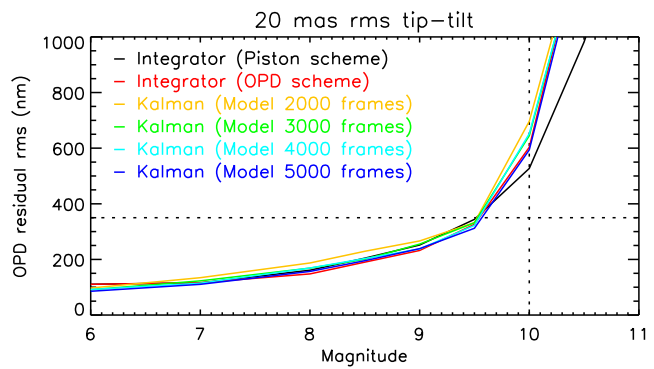}}
 \caption{OPD residuals as a function of magnitude, with flux variations due \refere{to 20~mas rms residual tip-tilt jitter}. The OPD disturbances are 10~$\mu$m rms atmospheric OPD and 150~nm rms vibration OPD \refere{(low vibration level)}. \refere{The dashed lines correspond to the specification for the fringe tracker of GRAVITY which is to stabilize the fringes below 350~nm rms on a reference star of magnitude $K=10$.}}
 \label{simu_tilt_levels}
\end{figure}

\begin{figure}
\resizebox{\hsize}{!}{\includegraphics{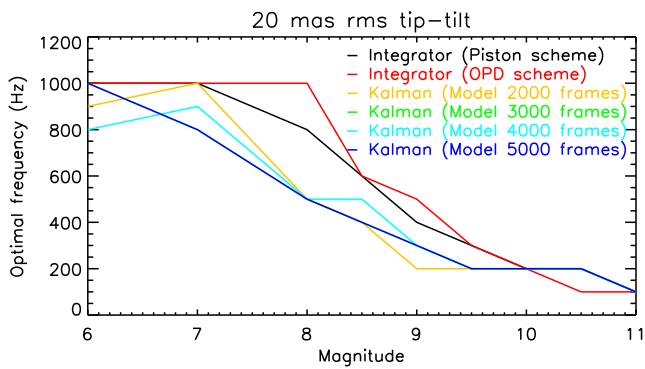}}
 \caption{Optimal frequencies as a function of the magnitude for conditions related to Fig.~\ref{simu_tilt_levels}.}
 \label{simu_tilt_levels_freq}
\end{figure}

\refere{We analyzed} the robustness of the controllers to flux dropouts by simulating a residual tip-tilt jitter of the beams with two different levels: \refere{the nominal level of} 15~mas rms, expected for GRAVITY after guiding correction, and \refere{a high tip-tilt level of} 20~mas rms. The residual OPDs at optimal loop frequency are presented in Fig.~\ref{simu_tilt_levels} as a function of magnitude \refere{for 20~mas rms} tip-tilt variations, with OPD vibrations of 150~nm rms and atmospheric disturbance of 10~$\mu$m rms. The corresponding optimal frequencies are presented in Fig.~\ref{simu_tilt_levels_freq}.

At high S/N (up to magnitude 8), the Kalman and the integrator controllers are equally insensitive to flux variations, and the OPD residuals are hardly larger with 20~mas than with 15~mas rms residual tip-tilt \refere{(compared with Fig.~\ref{simu_vib_levels}, middle panel). Actually in this regime, the star magnitude is far from the sensitivity limit of the instrument, and flux variations only result in variations in the precision of the OPD which are compensated by the weighted OPD combination provided by the redundant architecture of the interferometer (see Secs. \ref{noise_est} and~\ref{controllers})}. 

At low S/N, both controllers are almost equally robust to flux dropouts. Their performances as a function of magnitude are shifted of a quarter of magnitude between the two simulated tip-tilt levels. With a \refere{faint} reference star \refere{close to the sensitivity limit of the instrument}, flux variations induce frequent flux dropouts. The efficiency of the Kalman controller depends on the quality of the disturbance model, and consequently on the integrator controller accuracy. The predictive power of the Kalman filter is thus inhibited by the poor quality of the model at low S/N, and the Kalman controller is not more robust than the integrator controller.

\subsection{Performances at different atmospheric OPD levels\label{sec_simu_opd}}

\begin{figure}
 \resizebox{\hsize}{!}{\includegraphics{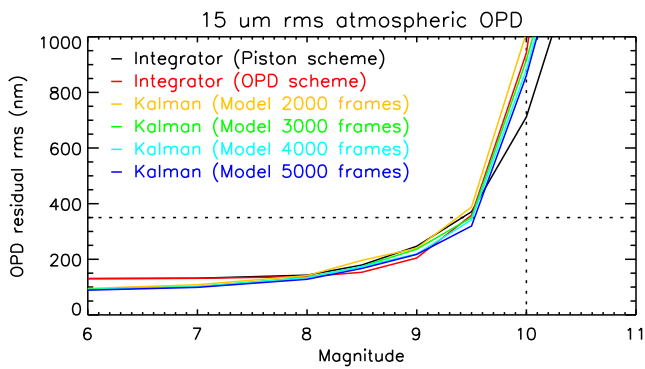}}
 \caption{OPD residuals as a function of magnitude \refere{with an atmospheric OPD of 15~$\mu$m rms}. The vibration level is 150~nm rms per baseline \refere{(low vibration level)}, and the flux variations are due to 15~mas rms beam tip-tilt variations.\refere{The dashed lines correspond to the specification for the fringe tracker of GRAVITY which is to stabilize the fringes below 350~nm rms on a reference star of magnitude $K=10$.}}
 \label{simu_opd_levels}
\end{figure}

\begin{figure}
\resizebox{\hsize}{!}{\includegraphics{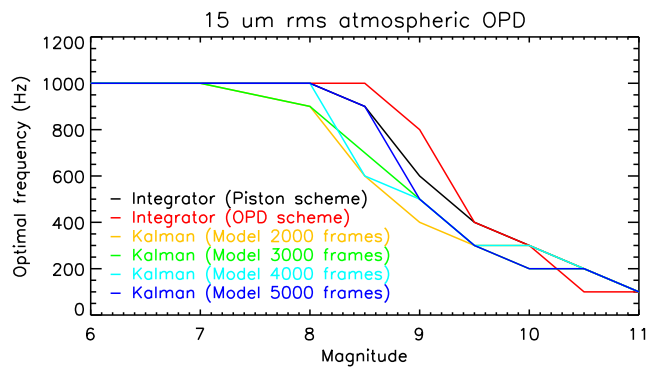}}
 \caption{Optimal frequencies as a function of the magnitude for conditions related to Fig.~\ref{simu_opd_levels}.}
 \label{simu_opd_levels_freq}
\end{figure}

\refere{We also analyzed the robustness of the controllers to a stronger atmospheric turbulence. In addition to the nominal level of 10~$\mu$m rms atmospheric OPD, we then simulated atmospheric turbulence of} 15~$\mu$m rms. \refere{It corresponds} respectively  to 1~\arcsec and 1.7~\arcsec seeing, assuming a Von K\'arm\'an model with 100~m outer scale \citep{Conan2000}. The residual OPDs \refere{for 15~$\mu$m atmospheric turbulence} are presented in Fig.~\ref{simu_opd_levels}, with 150~nm rms vibrations and 15~mas rms residual tip-tilt of the beams. The corresponding optimal frequencies are presented in Fig.~\ref{simu_opd_levels_freq}.

The performances of each controller as a function of magnitude are noticeably similar for both disturbance levels \refere{(compared with Fig.~\ref{simu_vib_levels}, middle panel)}. \refere{With a bright reference star, the controllers are insensitive to the atmospheric piston, thanks to the system redundancy which improve the OPD estimations (see Secs. \ref{noise_est} and~\ref{controllers}). The impact of a stronger atmospheric turbulence is to lower the sensitivity limit of the fringe tracker by half a magnitude for all controllers. The Kalman controller is not significantly more robust than the integrator controller for these observing conditions.}

\subsection{Kalman model identification on a bright star\label{sec_brightStar}}

\refere{From the results presented above, we conclude that the performance of the Kalman controller are mainly limited in the low S/N regime by inaccuracies in the identification of the model of the vibrations, due to a poor fringe stabilization during the POL sequences by the integrator controller. To overcome this limitation, we investigated the possibility of tracking the fringes on a bright star during the POL sequence to improve the disturbance model identification and the Kalman controller performance on the faint reference star.}

\refere{We thus performed additional simulations in which the POL sequences are acquired on a bright star of magnitude $K=7$ with the piston scheme integrator controller. Assuming that the bright star and the faint reference star are close to each other (within a few degrees) and that the telescope slew is fast enough, the light coming from the two stars is subject to atmospheric turbulence with similar statistical properties, and similar vibrations with telescopes pointing in very close directions, so the same disturbance model can be used. Only the Kalman gain matrix (Eq.~\ref{eqGainKal}) has to be adapted to optimally track the fringes on the faint reference star. For this we computed the Kalman gain with OPD uncertainties estimated from POL sequences simulated with the faint reference star.}

\begin{figure}
 \resizebox{\hsize}{!}{\includegraphics{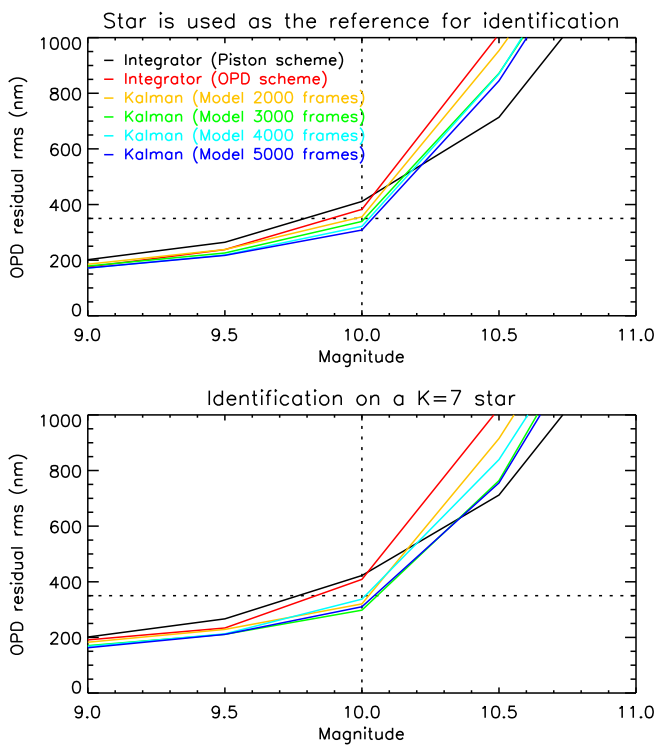}}
 \caption{OPD residuals as a function of magnitude, for 10~$\mu$m rms atmospheric OPD, and 15~mas rms residual tip-tilt \refere{variations, and the low vibration level (150~nm rms vibration OPD)}. Top: disturbance model is computed from a POL sequence acquired on the star itself \refere{(note that it simply a zoom-in of Fig.~\ref{simu_vib_levels}, middle panel.)}. Bottom: the model is computed observing a bright object ($K=7$). \refere{The dashed lines correspond to the specification for the fringe tracker of GRAVITY which is to stabilize the fringes below 350~nm rms on a reference star of magnitude $K=10$.}}
 \label{simu_idk7_expvib}
\end{figure}

\begin{figure}
 \resizebox{\hsize}{!}{\includegraphics{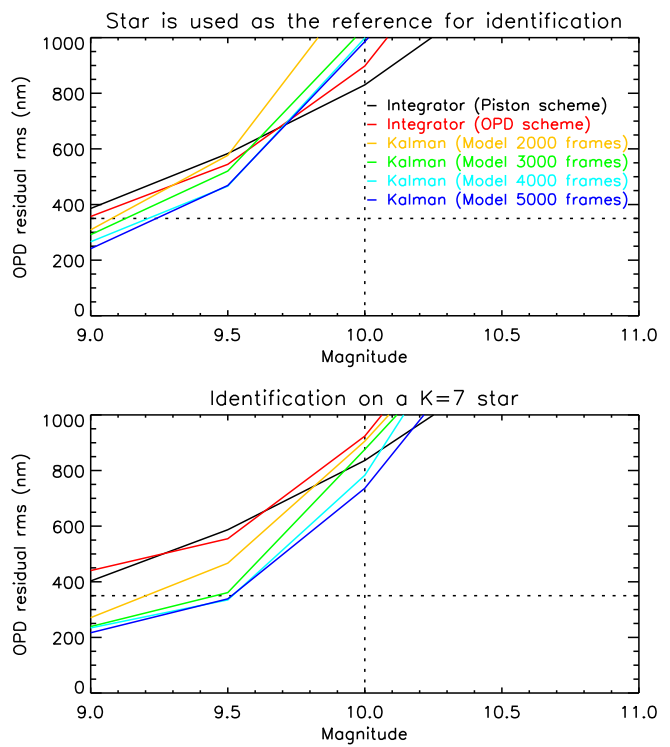}}
\caption{Same as Fig.~\ref{simu_idk7_expvib}, except \refere{that the simulations were done for the high vibration level (240 to 380~nm rms OPD)}. \refere{The top panel is here a zoom-in of  Fig.~\ref{simu_vib_levels}, bottom pannel.)}}
 \label{simu_idk7_curvib}
\end{figure}

\begin{figure}
 \resizebox{\hsize}{!}{\includegraphics{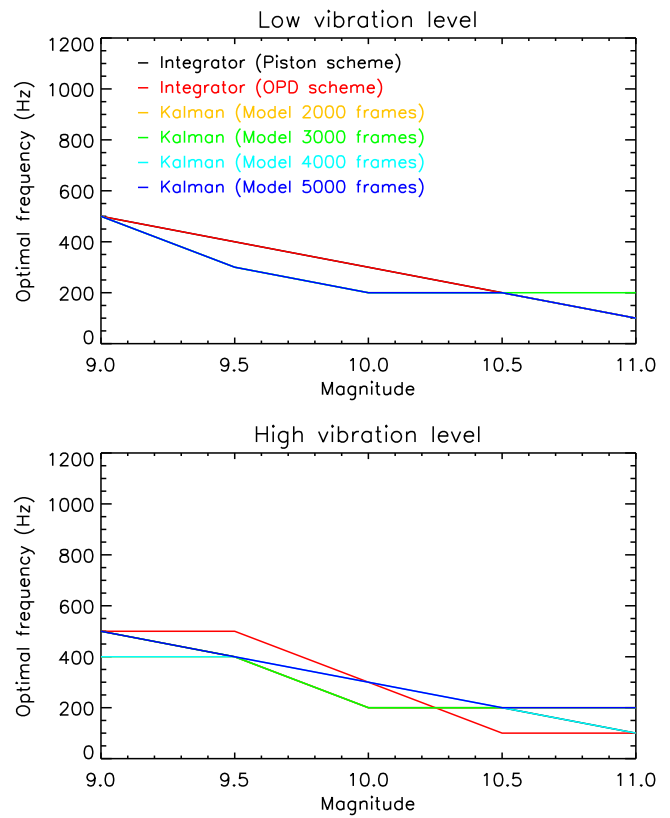}}
 \caption{Optimal frequencies as a function of the magnitude for conditions \refere{when the disturbance model is identified on a star of magnitude $K=7$.  Top panel is related to Fig.~\ref{simu_idk7_expvib} and bottom panel is related to Fig.~\ref{simu_idk7_curvib}}.}
 \label{simu_idk7_freq}
\end{figure}

\refere{For this analysis, we only ran simulations in the mid- and low-S/N regime, with reference star magnitudes ranging from 9 to 11 and loop frequencies from 100 to 500~Hz. The OPD residuals obtained with 10~$\mu$m rms atmospheric turbulence and 15~mas rms tip-tilt variations are presented in Figs.~\ref{simu_idk7_expvib} and \ref{simu_idk7_curvib} for the low and high vibration levels, respectively. The corresponding optimal loop frequencies are presented in Fig.~\ref{simu_idk7_freq}, top and bottom panels respectively.}

For the low vibration level (Fig.~\ref{simu_idk7_expvib}), \refere{there is slight improvement in the performance of the Kalman controller when identifying the disturbance model on high-S/N POL sequence, in particular for magnitude greater than 10. For stars brighter than $K=10$, the performances are mostly limited by the measurement noise level and  by the flux dropouts rather than by the accuracy of the model, and the gain in the OPD stabilization is of a few tens of nanometers only.}

For stronger vibrations (Fig.~\ref{simu_idk7_curvib}), the identification of the disturbances with high-S/N POL sequences clearly improves the Kalman controller performances. At magnitude 10 the OPD residuals are lowered by 250~nm rms by identifying the disturbances model with a $K=7$ star. Actually, at low S/N, the vibration level is too high for the integrator to properly track the fringes during the POL sequence (OPD residuals of 820~nm rms at magnitude 10), and the disturbance model estimated from these measurements is not accurate enough to provide an efficient fringe tracking with the Kalman controller. On the other hand, with \refere{a model identified from} high-S/N POL sequences, the \refere{control with the Kalman filter is then only limited by the flux dropouts.}

\section{Discussion\label{discussion}}

\refere{In this section we discuss the limitations of our simulations. We also analyze these results for the particular case of GRAVITY which we used as framework for our simulations, and we discuss their impact for the objectives of the instrument.}

\subsection{Limits on the simulations}

\refere{First of all, our simulations of the atmospheric turbulence are based on a Von K\`arm\`an model and parameters qualitatively chosen to generate disturbance sequences and spectra similar to those measured at VLTI. This model significantly differs from the one used at VLT to describe median observing conditions, based on a Kolmogorov model. According to this model, our simulations might be too optimistic when simulating 10~$\mu$m rms OPD, whereas 15~$\mu$m rms OPD would correspond to the median conditions (median atmospheric piston is characterized at VLTI by 310~nm rms OPD variations over a timescale of 48~ms when using a Kolmogorov model, which is the level we obtain by simulating a total turbulence level of 15~$\mu$m rms with the Von K\`arm\`an model). However, the metric defining the median atmospheric conditions might differ for the two models since the Kolmogorov model diverges at zero frequency and leads to a total piston variation depending on the time-scale, unlike a Von K\`arm\`an model.}

\refere{The second limitation of our simulation is that} we simulated an instantaneous response for the piston actuators, \refere{which is not realistic when working at high loop frequencies, and so for bright reference stars in the high S/N regime. Limited actuator bandwidths will reduce the absolute performance of all controllers but should not change the conclusion on the relative performance of the Kalman controller with respect to the integrator. In addition, the algorithm of the Kalman controller can be adapted to limited actuator bandwidths, as described by \conan{\citet{Correia2010}}, to improve the controller performance. In the specific case of GRAVITY, the piston actuators of the fringe tracker will have} a 3~dB bandwidth of 220~Hz. We thus anticipate a small loss in performances in fringe tracking on reference sources brighter than $K=8.5$ whose optimal frequency is greater than 500~Hz.

Finally, we considered time-invariant disturbances in our simulations. However, the atmospheric seeing slowly varies during on time-scales of a few minutes, and realistic instrumental vibrations may have varying frequency or amplitude, \refere{changing with, for instance, the wind speed or orientation, or the telescopes pointing}. \refere{If these variations are very significant, the disturbance model used for the Kalman filter have to be regularly updated on time-scales of 5-10~s to efficiently correct vibrations. For this, POL sequences can be reconstructed directly from tracking sequences using the Kalman controller, and the updated disturbance model can be computed on a dedicated computer without stopping the control loop. Since the Kalman controller is better at stabilizing fringes than a classical controller, the new POL sequences will have a better precision than the initial one obtained with the integrator controller, and disturbance model will not only be updated for the vibrations, but also be more accurate for the atmospheric turbulence. This strategy may be particularly valuable if the initial model has been identified by tracking the fringes on a bright star before Kalman-tracking on the faint reference star (see Sec.~\ref{sec_brightStar}): updating the initial model with new POL sequences acquired on the faint reference star would actually remove potential biases from the small angular separation between the bright initial star and the tracking reference star, if the atmospheric turbulence is not statistically spatio-temporally invariant.} 

\subsection{\refere{Performances with respect to} GRAVITY}

\refere{To enable the observation of the Galactic Center, GRAVITY will use IRS16C as the reference for the fringe tracker, a star of magnitude $K=9.7$. To achieve its science objectives, the fringe tracker is required to be able to track fringes} 
down to 350~nm rms on a $K=10$ reference source with the UTs, assuming vibrations below 150~nm rms OPD, 15~mas rms residual tip-tilt per beam, and median seeing conditions. Moreover, fringe tracking down to 300~nm rms OPD must be achieved without vibrations in the same observing conditions, assuming that OPD vibrations will add incoherently to the residuals.

\begin{table}
  \caption{OPD residuals at magnitude 10, with 150~nm rms OPD vibrations, 10~$\mu$m rms atmospheric OPD, 15~mas rms residual tip-tilt per beam.} 
 \label{table_gravity_spec} 
 \centering 
 \begin{tabular}{c c c c c c c} 
  \hline\hline 
  Controller & OPD residuals \\
	  & (nm rms)\\
\hline 
  Piston scheme integrator& 411\\
  OPD scheme integrator& 383\\
  Kalman (POL 2\,000 frames)& 356\\
  Kalman (POL 3\,000 frames)& 339 \\
  Kalman (POL 4\,000 frames)& 321\\
  Kalman (POL 5\,000 frames)& 308\\
  \hline 
 \end{tabular}
\end{table}

\begin{table}
  \caption{OPD residuals at magnitude 10, without longitudinal vibrations, 10~$\mu$m rms atmospheric OPD, 15~mas rms residual tip-tilt per beam.} 
 \label{table_gravity_spec_2} 
 \centering 
 \begin{tabular}{c c c c c c c} 
  \hline\hline 
  Controller 	& OPD residuals \\
		& (nm rms)\\
\hline 
  Piston scheme integrator& 279\\
  OPD scheme integrator& 366\\
  Kalman (POL 2\,000 frames)& 235\\
  Kalman (POL 3\,000 frames)& 239 \\
  Kalman (POL 4\,000 frames)& 230\\
  Kalman (POL 5\,000 frames)& 228\\
  \hline 
 \end{tabular}
\end{table}
Our simulations \refere{show that a} Kalman controller \refere{will} enable the fringe tracker of GRAVITY to \refere{meet these requirements, but not an} integrator controller. The OPD residuals for each controller for fringe tracking at magnitude 10 under the specified conditions are detailed in table~\ref{table_gravity_spec}. The use of a long POL sequence of 5\,000 frames to identify the disturbance model provides a 40~nm rms margin to the specified OPD residual level, at the cost of memory and computation time to identify the model. 

Without vibrations, the Kalman algorithm also provides a better control than the integrator controller, although both algorithms reach the specification of \refere{the fringe tracker of} GRAVITY. The corresponding OPD residuals are detailed in table~\ref{table_gravity_spec_2}. The Kalman controller provides $\sim$70~nm rms margin to the specified OPD residual level.

However, these performances at magnitude 10 are close to the limits of the Kalman controller. They quickly degrade with stronger disturbances: the fringe tracker sensitivity is lowered by half a magnitude for 20~mas rms tip-tilt of the beams or for 15~$\mu$m rms atmospheric OPD. \refere{More efforts to} damp the vibration level at VLTI is also crucial \refere{for GRAVITY to reach its science objectives}: with the current vibration level on the UTs, fringe tracking at magnitude 10 can not be expected to provide OPD residuals lower than 800~nm rms, and the 350~nm rms residual level can only be reached with a $K=9.2$ star.

 \section{Conclusion\label{conc}}
\refere{In this paper, we studied the performances of two different algorithms to track fringes with an interferometer.}
We performed simulations of fringe tracking with different reference star magnitude, with realistic disturbances based on observing conditions at VLTI (flux variations, vibrations, and seeing conditions), and compared the performances obtained with an optimized integrator controller and with a controller based on a Kalman filter, \refere{which can predict statistically optimal commands based on a model of the disturbance}. \refere{We based our simulations on the architecture and design of the GRAVITY instrument, four-telescope combiner that will be installed at VLTI end of 2014.}

\refere{In the high S/N regime, we found that the efficiency to track fringes is globally constant for each controller. Their performance is limited by the maximal loop frequency that the system can achieve (1~kHz in our simulations), and thus by the controller bandwidths.}
\refere{In presence of vibrations,} the Kalman controller is significantly more efficient than an integrator controller \refere{at high S/N}. Whereas the integrator controller is clearly limited by the vibrations, the Kalman \refere{controller} is almost insensitive to the vibration level, and more generally to the disturbance level in this range of magnitude. The disturbance model is accurately estimated and the actuator commands are computed with accurate prediction of the vibrations. 

At lower S/N, the performances of both controllers significantly decrease with the star magnitude. \refere{For the Kalman controller, supposed to be able to predict optimal commands,} it can be attributed to an inaccurate model of the disturbances. Because of poor-efficiency fringe tracking with the integrator controller during the preliminary sequence used to calibrate the disturbances, the model of disturbance used by the Kalman controller is indeed not accurately estimated and the corrections are not properly adapted. 
For observing conditions leading to OPD residuals of more than $\sim\lambda/4$ with the integrator controller (faint magnitude, high disturbance or flux dropout level), the noise on the preliminary sequence is too important to provide an accurate identification of the disturbance model and the Kalman controller is less robust than a classical integrator controller.

We also found that in this situation, the piston control can be improved by identifying the disturbance model from high S/N preliminary sequences using a bright star with the integrator controller before tracking the fringes on the faint reference star with the Kalman controller. 

\refere{Finally, our results show that the VLTI instrument GRAVITY will be able to reach its science objectives with a fringe tracker using a Kalman controller, by stabilizing the fringes down to 350~nm rms with a reference star of magnitude $K=10$ under median atmospheric conditions. However, this performance depends on a significant effort to mitigate vibration at VLTI down to 150~nm rms OPD on the UTs. Four-telescope fringe tracking is to be demonstrated with both algorithms in laboratory during the integration of GRAVITY in mid-2014, then on sky by 2015.}

\begin{acknowledgements}
\conan{We are grateful to Cyril Petit  and Serge Meimon for initial discussions and advices on Kalman/LQG control, and to Julien Lozi for several input discussions on practical implementation of the Kalman algorithm. We also thank Jean-Marc Conan, Caroline Kulcsar and Henri-François Raynaud for substantial suggestions on state-of-the-art of Kalman/LQG command for adaptive optics.}
EC is grateful to DGA and CNRS/INSU for PhD fellowship grant.
\end{acknowledgements}
\bibliographystyle{aa} 
\bibliography{biblio-gravity-noabb.bib}

\appendix
\input{Appendixes}

\end{document}

%% file: Appendixes.tex
\section{Derivation of the noise of the phase \label{phi_noise_app}}

\begin{figure}
\resizebox{\hsize}{!}{\includegraphics{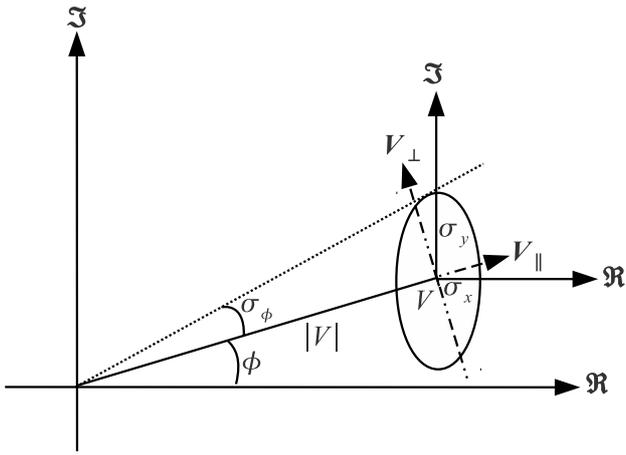}}
 \caption{Representation of the complex visibility $V$ and of the noise $\sigma_V$in the complex plane $(V, \vec{\Re},\vec{\Im})$ and in the radial basis $(V, \vec{V_\parallel},\vec{V_\perp})$.}
 \label{phi_noise}
\end{figure}

Let $V$ be a complex visibility of phase and modulus $\phi$ and $|V|$ respectively in the complex plane.
Let $\sigma_\mathrm{V}=(\sigma_\mathrm{x}, \sigma_\mathrm{y})$ be the noise on $V$ in the complex plane, with $\sigma_\mathrm{x}$ and $\sigma_\mathrm{y}$ the real and imaginary parts respectively. Assuming that the noise are not correlated, the noise on the complex vector $V$ is thus described by an ellipse of semi-minor axis $\sigma_\mathrm{x}$ and semi-major axis $\sigma_\mathrm{y}$ in the basis $\left(V,\vec{\Im},\vec{\Re}\right)$ centered on $V$ (see Fig. \ref{phi_noise}). A point on the ellipse with coordinates $(x,y)$ in this basis verifies the equation :
\begin{equation}
 \left(\frac{x}{\sigma_\mathrm{x}}\right)^2+\left(\frac{y}{\sigma_\mathrm{y}}\right)^2 =1
\end{equation}
In the radial orthonormal basis $(V,\vec{V_\mathrm{\parallel}},\vec{V_\mathrm{\perp}})$ defined respectively by the axis parallel and orthogonal to $\vec{V}$, the same point of coordinates $(u,v)$ verifies the equation:
\begin{equation}
 \left(\frac{u\cos\phi-v\sin\phi}{\sigma_\mathrm{x}}\right)^2+\left(\frac{u\sin\phi+v\cos\phi}{\sigma_\mathrm{y}}\right)^2 =1
\end{equation}

The noise $\sigma_\phi$ on the phase can be approximated by the differential phase between the $V$ and the point of the ellipse whose partial derivative in $u$ is null (which corresponds to the point of maximal coordinate on the axis $\vec{B_\perp}$). Two points verify these relations, with coordinates $(u_1,v_1)$ and $(u_2,v_2)$ in the basis $(V,\vec{V_\mathrm{\parallel}},\vec{V_\mathrm{\perp}})$:
\begin{equation}
 u_{1,2}=\pm \frac{\cos\phi\sin\phi(\sigma_\mathrm{y}^2-\sigma_\mathrm{x}^2)}{\sqrt{\sigma_\mathrm{y}^2\cos\phi^2+\sigma_\mathrm{x}^2\sin\phi^2}}
\end{equation}
\begin{equation}
 v_{1,2}=\pm \sqrt{\sigma_\mathrm{y}^2\cos\phi^2+\sigma_\mathrm{x}^2\sin\phi^2}
\end{equation}
The noise $\sigma_\phi$ on the phase is thus:
\begin{equation}
 \sigma_\phi= \max\left(\left|\arctan\left(\frac{v_1}{|V|+u_1}\right)\right|, \left|\arctan\left(\frac{v_2}{|V|+u_2}\right)\right|\right)
\end{equation}

This expression neglect the covariance terms between the real and the imaginary part of the complex vector. We refer to \citet{Williams2006} for the analytical expression on the noise on the phase in case of correlation between both measurements.

\section{Derivation of the noise of the cross product operator \label{Noise_Xp}}

The cross-product operator for a single baseline can be expressed as a function $z$ of two complex numbers $x$ and $y$:
\begin{equation}
 z=xy^*
\end{equation}
The expression of the real and imaginary parts of $z$ are developed respectively as:
\begin{equation}
 \Re(z)= \Re(x)\Re(y)+\Im(x)\Im(y)
\end{equation}
\begin{equation}
 \Im(z)=\Re(y)\Im(x)-\Re(x)\Im(y)
\end{equation}
The variance of these expressions is expressed respectively as:
\begin{equation}
 \sigma_{\Re(z)}^2= \Re(y)^2\sigma_{\Re(x)}^2+\Re(x)^2\sigma_{\Re(y)}^2 +  \Im(y)^2\sigma_{\Im(x)}^2 +\Im(y)^p2\sigma_{\Im(y)}^2
\end{equation}
\begin{equation}
 \sigma_{\Im(z)}^2=\Im(y)^2\sigma_{\Re(x)}^2+\Im(x)^2\sigma_{\Re(y)}^2 +  \Re(y)^2\sigma_{\Im(x)}^2 +\Re(x)^2\sigma_{\Im(y)}^2 				             	
\end{equation}
With $\sigma_{\Re(x)}^2$ and $\sigma_{\Im(x)}^2$ the variance of the real and imaginary part of $x$ respectively.